\input ./style/arxiv-general.cfg
\documentclass[aos,MSNbibl,seceqn,nameyear,dvips]{arximspdf}
\makeatletter
   \@ifpackageloaded{graphicx}{}{\usepackage{graphicx}}
\makeatother
\usepackage{dcolumn}

\doi{10.1214/15-AOS1364}
\volume{44}
\issue{1}
\pubyear{2016}
\firstpage{219}
\lastpage{254}
\docsubty{FLA}

\makeatletter
\newcolumntype{d}[1]{D{.}{.}{#1}}

\newcommand{\eqref}[1]{(\ref{#1})}
\newtheorem{thmm}{Theorem}[section]
\newproclaim{defn}{Definition}[section]
\newtheorem{lem}{Lemma}[section]
\newtheorem{cor}{Corollary}[section]

\newproclaim{assum}{Assumption}[section]
\newproclaim{exm}{Example}[section]
\newproclaim{remark}{Remark}[section]
\renewcommand{\hat}{\widehat}

\def\bA{\mathbf A}
\def\bB{\mathbf B}
\def\bD{\mathbf D}
\def\bE{\mathbf E}
\def\bff{\mathbf f}\def\bF{\mathbf F}
\def\bG{\mathbf G}
\def\bH{\mathbf H}
\def\bI{\mathbf I}
\def\bK{\mathbf K}
\def\bM{\mathbf M}
\def\bP{\mathbf P}
\def\bR{\mathbf R}
\def\bS{\mathbf S}
\def\bU{\mathbf U}
\def\bV{\mathbf V}
\def\bW{\mathbf W}
\def\bX{\mathbf X}
\def\bY{\mathbf Y}

\def\bbeta{\bolds{\beta}}

\def\bGamma{\bolds{\Gamma}}
\def\bDelta{\bolds{\Delta}}
\def\bSigma{\bolds{\Sigma}}
\def\bLambda{\bolds{\Lambda}}

\renewcommand{\hat}{\widehat}

\newcommand{\diag}{\operatorname{diag}}
\newcommand{\cov}{\operatorname{cov}}
\newcommand{\tr}{\operatorname{tr}}

\def\var{\operatorname{var}}

\newcommand{\bchix}{\bolds{\mathcal{X}}}
\newcommand{\bone}{\mathbf{1}}
\newcommand{\bzero}{\mathbf{0}}
\newcommand{\bXi}{\bolds{\Xi}}
\newcommand{\bepsilon}{\bolds{\varepsilon}}

\newcommand{\hb}{\widehat{\mathbf b}}
\newcommand{\hF}{\widehat\bF}
\newcommand{\hG}{\widehat\bG}
\newcommand{\hXi}{\widehat\bXi}
\newcommand{\Sig}{\bolds{\Sigma}}
\newcommand{\vecc}{\mathrm{vec}}

\makeatother

\begin{document}
\begin{frontmatter}

\title{Projected principal component analysis in factor~models\thanksref{T1}}
\runtitle{Projected-PCA}

\begin{aug}
\author[A]{\fnms{Jianqing}~\snm{Fan}\thanksref{M1}\ead[label=e1]{jqfan@princeton.edu}},
\author[B]{\fnms{Yuan}~\snm{Liao}\thanksref{M2}\ead[label=e2]{yuanliao@umd.edu}}
\and
\author[A]{\fnms{Weichen}~\snm{Wang}\corref{}\thanksref{M1}\ead[label=e3]{weichenw@princeton.edu}}
\runauthor{J. Fan, Y. Liao and W. Wang}
\affiliation{Princeton University\thanksmark{M1} and University of
Maryland\thanksmark{M2}}
\address[A]{J. Fan\\
W. Wang\\
Department of ORFE\\
Sherrerd Hall\\
Princeton University\\
Princeton, New Jersey 08544\\
USA\\
\printead{e1}\\
\phantom{E-mail:\ }\printead*{e3}}

\address[B]{Y. Liao\\
Department of Mathematics\\
University of Maryland\\
College Park, Maryland 20742\\
USA\\
\printead{e2}}
\thankstext{T1}{Supported by NSF Grants DMS-12-06464 and
2R01-GM072611-9, and the Research and Scholarship award from University
of Maryland.}
\end{aug}

%
\received{\smonth{1} \syear{2015}}
%
\revised{\smonth{7} \syear{2015}}

%
\begin{abstract}
This paper introduces a Projected Principal Component Analysis 
(Projected-PCA), which employs principal component analysis to the
projected (smoothed) data matrix onto a given linear space spanned by
covariates. When it applies to high-dimensional factor analysis, the
projection removes noise components. We show that the unobserved latent
factors can be more accurately estimated than the conventional PCA if
the projection is genuine, or more precisely, when the factor loading
matrices are related to the projected linear space. When the
dimensionality is large, the factors can be estimated accurately even
when the sample size is finite. We propose a flexible semiparametric
factor model, which decomposes the factor loading matrix into the
component that can be explained by subject-specific covariates and the
orthogonal residual component. The covariates' effects on the factor
loadings are further modeled by the additive model via sieve
approximations. By using the newly proposed Projected-PCA, the rates of
convergence of the smooth factor loading matrices are obtained, which
are much faster than those of the conventional factor analysis. The
convergence is achieved even when the sample size is finite and is
particularly appealing in the high-dimension-low-sample-size situation.
This leads us to developing nonparametric tests on whether observed
covariates have explaining powers on the loadings and whether they
fully explain the loadings. The proposed method is illustrated by both
simulated data and the returns of the components of the S\&P 500 index.
\end{abstract}

%
%
\end{frontmatter}

\section{Introduction}\label{sec1}



Factor analysis is one of the most useful tools for modeling common
dependence among multivariate outputs. Suppose that we observe data $\{
y_{it}\}_{i\leq p, t\leq T}$ that can be decomposed as
%
\begin{equation}
\label{eq1.1} y_{it}=\sum_{k=1}^K
\lambda_{ik}f_{tk}+u_{it},\qquad i=1,\ldots ,p, t=1,
\ldots,T,
\end{equation}
where $\{f_{t1},\ldots, f_{tK}\}$ are unobservable common factors; $\{
\lambda_{i1},\ldots,\lambda_{iK}\}$ are corresponding factor loadings
for variable $i$, and $u_{it}$ denotes the idiosyncratic component that
cannot be explained by the static common component. Here, $p$ and $T$,
respectively, denote the dimension and sample size of the data.

Model \eqref{eq1.1} has broad applications in the statistics
literature. For instance, $\mathbf y_t = (y_{1t}, \ldots, y_{pt})'$
can be
expression profiles or blood oxygenation level dependent (BOLD)
measurements for the $t${th} microarray, proteomic or fMRI-image,
whereas $i$ represents a gene or protein or a voxel. See, for example,
\citet
{leek2008general,friguet2009factor,efron2010correlated,desai2012cross,FHG8}.
The separations between the common factors and
idiosyncratic components are carried out by the low-rank plus sparsity
decomposition. See, for example, \citet
{candes2009exact,koltchinskii2011nuclear,negahban2011estimation,POET,cai2013sparse,ma2013sparse}.

The factor model (\ref{eq1.1}) has also been extensively studied in the
econometric literature, in which $\mathbf y_t $ is the vector of economic
outputs at time $t$ or excessive returns for individual assets on day
$t$. The unknown factors and loadings are typically estimated by the
principal component analysis (PCA) and the separations between the
common factors and idiosyncratic components are characterized via
{static pervasiveness} assumptions. See, for instance, \citet
{SW02,bai03,BN02,BT11,LamYao} among others.
In this paper, we consider static factor model, which differs from the
dynamic factor model [\citet{ForLip01}, \citeauthor{ForHalLipRei00}
(\citeyear{ForHalLipRei00,ForHalLipZaf15})].
The dynamic model allows more general infinite dimensional
representations. For this type of model, the frequency domain PCA
[\citet
{Bri81}] was applied on the spectral density. The so-called \textit
{dynamic pervasiveness} condition also plays a crucial role in
achieving consistent estimation of the spectral density.

Accurately estimating the loadings and unobserved factors are very
important in statistical applications. In calculating the
false-discovery proportion for large-scale hypothesis testing, one
needs to adjust accurately the common dependence via subtracting it
from the data in (\ref{eq1.1}) [\citet
{leek2008general,friguet2009factor, efron2010correlated,
desai2012cross, FHG8}]. In financial applications, we would like to
understand accurately how each individual stock depends on unobserved
common factors in order to appreciate its relative performance and risks.
In the aforementioned applications, dimensionality is much higher than
sample-size.
However, the existing asymptotic analysis shows that the consistent
estimation of the parameters in model (\ref{eq1.1}) requires a
relatively large $T$. In particular, the individual loadings can be
estimated no faster than $O_P(T^{-1/2})$. But large sample sizes are
not always available. Even with the availability of ``Big Data,''
heterogeneity and other issues make direct applications of (\ref
{eq1.1}) with large $T$ infeasible.
For instance, in financial applications, to pertain the stationarity in
model (\ref{eq1.1}) with time-invariant loading coefficients, a
relatively short time series is often used. To make observed data less
serially correlated, monthly returns are frequently used to reduce the
serial correlations, yet a monthly data over three consecutive years
contain merely 36 observations.

\subsection{This paper}\label{sec1.1}

To overcome the aforementioned problems, and when relevant covariates
are available, it may be helpful to incorporate them into the model.
Let $\bX_i=(X_{i1}, \ldots, X_{id})'$ be a vector of $d$-dimensional
covariates associated with the $i$th variables. In the seminal
papers by \citet{CL07} and \citet{CMO}, 
the authors studied the following semi-parametric factor model:
%
\begin{equation}
\label{eq1.2} y_{it}=\sum_{k=1}^K
g_k(\bX_{i})f_{tk}+u_{it},\qquad i=1,
\ldots,p, t=1,\ldots,T,
\end{equation}
where loading coefficients in (\ref{eq1.1}) are modeled as $\lambda
_{ik} = g_k (\bX_i)$ for some functions $g_k (\cdot)$. For instance, in
health studies, $\bX_i$ can be
individual characteristics (e.g., age, weight, clinical and genetic
information); in financial applications $\bX_i$ can be a vector of
firm-specific characteristics (market capitalization, price-earning
ratio, etc.).


The semiparametric model (\ref{eq1.2}), however, can be restrictive in
many cases, as it requires that the loading matrix be fully explained
by the covariates. A natural relaxation is the following semiparametric model:
%
\begin{equation}
\label{eq1.3} \lambda_{ik} = g_k(\bX_{i}) +
\gamma_{ik},\qquad i = 1, \ldots, p, k =1, \ldots, K,
\end{equation}
where $\gamma_{ik}$ is the component of loading coefficient that cannot
be explained by the covariates $\bX_i$. Let $\bolds{\gamma}_i =
(\gamma_{i1},
\ldots, \gamma_{iK})'$. We assume that $\{\bolds{\gamma}_i\}_{i\leq
p}$ have
mean zero, and are independent of $\{\bX_i\}_{i\leq p}$ and $\{u_{it}\}
_{i\leq p, t\leq T}$. In other words, we impose the following factor structure:
%
\begin{equation}
\label{eq1.4} y_{it}=\sum_{k=1}^K
\bigl\{g_k(\bX_{i})+\gamma_{ik} \bigr\}
f_{tk}+u_{it},\qquad i=1,\ldots,p, t=1,\ldots,T,
\end{equation}
which reduces to model (\ref{eq1.2}) when $\gamma_{ik} = 0$ and model
(\ref{eq1.1}) when $g_k(\cdot) = 0$. When $\bX_i$ genuinely explains a
part of loading coefficients $\lambda_{ik}$, the variability of
$\gamma
_{ik}$ is smaller than that of $\lambda_{ik}$. Hence, the coefficient
$\gamma_{ik}$ can be more accurately estimated by using regression
model (\ref{eq1.3}), as long as the functions $g_k(\cdot)$ can be
accurately estimated.

Let $\bY$ be the $p\times T$ matrix of $y_{it}$, $\bF$ be the
$T\times
K$ matrix of $f_{tk}$, $\bG(\bX)$ be the $p\times K$ matrix of
$g_k(\bX
_i)$, $\bGamma$ be the $p\times K$ matrix of $\gamma_{ik}$ and $\bU$ be
$p\times T$ matrix of $u_{it}$. Then model (\ref{eq1.4}) can be written
in a more compact matrix form:
%
\begin{equation}
\label{eq1.5} \bY=\bigl\{\bG(\bX)+\bGamma\bigr\} \bF' + \bU.
\end{equation}
We treat the loadings $\bG(\bX)$ and $\bGamma$ as realizations of
random matrices throughout the paper. This model is also closely
related to the \textit{supervised singular value decomposition} model,
recently studied by \citet{li2015supervised}. The authors showed that
the model is useful in studying the gene expression and
single-nucleotide polymorphism (SNP) data, and proposed an EM algorithm
for parameter estimation.

We propose a Projected-PCA estimator for both the loading functions and
factors. Our estimator is constructed by first projecting $\bY$ onto
the sieve space spanned by $\{\bX_i\}_{i\leq p}$, then applying PCA to
the projected data or fitted values. Due to the approximate
orthogonality condition of $\bX$, $\bU$ and $\bGamma$, the projection
of $\bY$ is approximately $\bG(\bX)\bF'$, as the smoothing projection
suppresses the noise terms $\bGamma$ and $\bU$ substantially.
Therefore, applying PCA to the projected data allows us to work
directly on the sample covariance of $\bG(\bX)\bF'$, which is $\bG
(\bX
)\bG(\bX)'$ under normalization conditions. This substantially improves
the estimation accuracy, and also facilitates the theoretical analysis.
In contrast, the
traditional PCA method for factor analysis [e.g., \citet{SW02}, \citet
{BN02}] is no longer suitable in the current context. Moreover, the
idea of Projected-PCA is also potentially applicable to dynamic factor
models of \citet{ForHalLipRei00}, by first projecting the data onto the
covariate space. 

The asymptotic properties of the proposed estimators are carefully studied.
We demonstrate that as long as the projection is genuine, the
consistency of the proposed estimator for latent factors and loading
matrices requires only $p\rightarrow\infty$, and $T$ does not need to
grow, which is attractive in the typical high-dimension-low-sample-size
(HDLSS) situations [e.g., \citet{JM09,shen2013consistency,ahn2007high}].
In addition, if both $p$ and $T$ grow simultaneously, then with
sufficiently smooth $g_k(\cdot)$, using the sieve approximation, the
rate of convergence for the estimators is much faster than those of the
existing results for model (\ref{eq1.1}). Typically, the loading
functions can be estimated at a convergence rate $O_P((pT)^{-1/2})$,
and the factor can be estimated at $O_P(p^{-1})$. Throughout the paper,
$K=\dim(\bff_t)$ and $d=\dim(\bX_i)$ are assumed to be constant and do
not grow.



Let $\bLambda$ be a $p\times K$ matrix of $(\lambda_{ik})_{T\times K}$.
Model (\ref{eq1.3}) implies a decomposition of the loading matrix:
\[
\bLambda=\bG(\bX)+\bGamma,\qquad E(\bGamma|\bX) = 0,
\]
where $\bG(\bX)$ and $\bGamma$ are orthogonal loading components in the
sense that $E
\bG(\bX) \bGamma' = 0$. We conduct two specification tests for the
hypotheses:
\[
H_0^1: \bG(\bX)=0 \qquad \mbox{a.s.}\quad \mbox{and}\quad
H_0^2: \bGamma =0 \qquad \mbox{a.s.}
\]
%
The first problem is about testing whether the observed covariates have
explaining power on the loadings. If the null hypothesis is rejected,
it gives us the theoretical basis to employ the Projected-PCA, as the
projection is now genuine. Our empirical study on the asset returns
shows that firm market characteristics do have explanatory power on the
factor loadings, which lends further support to our Projected-PCA
method. The second tests whether covariates fully explain the loadings.
Our aforementioned empirical study also shows that model (\ref{eq1.2})
used in the financial econometrics literature is inadequate and more
generalized model (\ref{eq1.5}) is necessary. As claimed earlier, even
if $H_0^2$ does not hold, as long as $\bG(\bX)\neq0$, the
Projected-PCA can still consistently estimate the factors as
$p\rightarrow\infty$, and $T$ may or may not grow.
Our simulated experiments confirm that the estimation accuracy is
gained more significantly for small $T$'s.
This shows one of the benefits of using our Projected-PCA method over
the traditional methods in the literature.


In addition, as a further illustration of the benefits of using
projected data, we apply the Projected-PCA to consistently estimate the
number of factors, which is similar to those in \citet{AH} and \citet
{LamYao}. Different from these authors, our method applies to the
projected data, and we demonstrate numerically that this can
significantly improve the estimation accuracy.

We focus on the case when the observed covariates are time-invariant.
When $T$ is small, these covariates are approximately locally constant,
so this assumption is reasonable in practice. On the other hand, there
may exist individual characteristics that are time-variant [e.g., see
\citet{park}]. We expect the conclusions in the current paper to still
hold if some smoothness assumptions are added for the time varying
components of the covariates. Due to the space limit, we provide
heuristic discussions on this case in the supplementary material of
this paper [\citet{PPCAsupp}]. In addition, note that in the usual factor
model, $\bLambda$ was assumed to be deterministic. In this paper,
however, $\bLambda$ is mainly treated to be stochastic, and potentially
depend on a set of covariates. But we would like to emphasize that the
results presented in Section~\ref{1541512515} under the framework of more general
factor models hold regardless of whether $\bLambda$ is stochastic or
deterministic. Finally, while some financial applications are presented
in this paper, the Projected-PCA is expected to be useful in broad
areas of statistical applications [e.g., see \citet{li2015supervised}
for applications in gene expression data analysis].

\subsection{Notation and organization}\label{sec1.2}

Throughout this paper, for a matrix $\bA$, let $\|\bA\|_F=\tr
^{1/2}(\bA
'\bA)$ and $\|\bA\|_2=\lambda_{\max}^{1/2}(\bA'\bA)$, $\|\bA\|
_{\max
}=\max_{ij}|A_{ij}|$ denote its Frobenius, spectral and max- norms. Let
$\lambda_{\min}(\cdot)$ and $\lambda_{\max}(\cdot)$ denote the minimum
and maximum eigenvalues of a square matrix. For a vector $\mathbf v$,
let $\|
\mathbf v\|$ denote its Euclidean norm.


The rest of the paper is organized as follows. Section~\ref{sec2} introduces the
new Projected-PCA method and defines the corresponding estimators for
the loadings and factors. Sections~\ref{1541512515} and \ref{s4} provide asymptotic analysis
of the introduced estimators. Section~\ref{sec5} introduces new specification
tests for the orthogonal decomposition of the semiparametric loadings.
Section~\ref{sec6} concerns about estimating the number of factors. Section~\ref{sec7}
presents numerical results. Finally, Section~\ref{sec8} concludes. All the
proofs are given in the \hyperref[app]{Appendix} and the supplementary material [\citet{PPCAsupp}].

\section{Projected principal component analysis}\label{sec2}

\subsection{Overview}\label{sec2.1}

In the high-dimensional factor model, let $\bLambda$ be the $p\times K$
matrix of loadings. Then the general model (\ref{eq1.1}) can be
written as
%
\begin{equation}
\label{eq2.1} \bY=\bLambda\bF'+\bU.
\end{equation}

Suppose we additionally observe a set of covariates $\{\bX_{i}\}
_{i\leq p}$. The basic idea of the Projected-PCA is to smooth the
observations $\{Y_{it}\}_{i\leq p}$ for each given day $t$ against its
associated covariates. More specifically, let $\{\hat{Y}_{it}\}_{i\leq
p}$ be the fitted value after regressing $\{Y_{it}\}_{i\leq p}$ on $\{
\bX_{i}\}_{i\leq p}$ for each given $t$. This results in a smooth or
projected observation matrix $\hat{\bY}$, which will also be denoted by
$\bP\bY$. The Projected-PCA then estimates the factors and loadings by
running the PCA based on the projected data $\hat{\bY}$. 

Here, we heuristically describe the idea of Projected-PCA; rigorous
analysis will be carried out afterward. Let $\bchix$ be a space spanned
by $\bX=\{\bX_i\}_{i\leq p}$, which is orthogonal to the error matrix
$\bU$. Let $\bP$ denote the projection matrix onto $\bchix$ [whose
formal definition will be given in (\ref{eq2.5}) below. At the
population level, $\bP$ approximates the conditional expectation
operator $E(\cdot|\bX)$, which satisfies $E(\bU|\bX)=0$], then $\bP
^2=\bP$ and $\bP\bU\approx0$. Hence, analyzing the projected data
$\hat
{\bY}=\bP\bY$ is an approximately noiseless problem, and the sample
covariance has the following approximation:
%
\begin{equation}
\label{eq2.1a} \frac{1}{T} \hat{\bY}' \hat{\bY} =
\frac{1}{T} \bY' \bP\bY \approx \frac{1}{T} \bF
\bLambda' \bP\bLambda\bF'.
\end{equation}

We now argue that $\bF$ and $\bP\bLambda$ can be recovered from the
projected data $\hat{\bY}$ under some suitable normalization condition.
The normalization conditions we impose are
%
\begin{equation}
\label{eq2.2} \frac{1}{T}\bF'\bF=\bI_K,\qquad
\bLambda'\bP\bLambda\mbox{ is a diagonal matrix with distinct
entries.}\hspace*{-20pt}
\end{equation}
%
Under this normalization, using \eqref{eq2.1a}, $ \frac{1}{T}\bY'\bP
\bY
\bF\approx\bF\bLambda'\bP\bLambda$. We conclude that the columns
of $\bF
$ are approximately $\sqrt{T}$ times the first $K$ eigenvectors of the
$T\times T$ matrix $\frac{1}{T}\bY'\bP\bY$.
Therefore, the Projected-PCA naturally defines a factor estimator
$\widehat\bF$ using the first $K$ principal components of $\frac
{1}{T}\bY'\bP\bY$.

The projected loading matrix $\bP\bLambda$ can also be recovered from
the projected data $\bP\bY$ in two (equivalent) ways. Given $\bF$, from
$ \frac{1}{T}\bP\bY\bF= \bP\bLambda+\frac{1}{T}\bP\bU\bF$, we
see $\bP
\bLambda\approx\frac{1}{T}\bP\bY\bF$. Alternatively, consider the
$p\times p$ projected sample covariance:
\[
\frac{1}{T}\bP\bY\bY'\bP=\bP\bLambda\bLambda'
\bP+\widetilde \bDelta,
\]
where $\widetilde\bDelta$ is a remaining term depending on $\bP\bU
$. %
Right multiplying $\bP\bLambda$ and ignoring $\widetilde\bDelta$, we
obtain $(\frac{1}{T}\bP\bY\bY'\bP)\bP\bLambda\approx\bP
\bLambda(\bLambda
'\bP\bLambda)$. Hence, the (normalized) columns of $\bP\bLambda$
approximate the first $K$ eigenvectors of $\frac{1}{T}\bP\bY\bY'\bP$,
the $p\times p$ sample covariance matrix based on the projected data.
Therefore, we can either estimate $\bP\bLambda$ by $\frac{1}{T}\bP
\bY
\widehat\bF$ given $\widehat\bF$, or by the leading eigenvectors of
$\frac{1}{T}\bP\bY\bY'\bP$. In fact, we shall see later that these two
estimators are equivalent. If in addition, $\bLambda=\bP\bLambda$, that
is, the loading matrix belongs to the space $\bchix$, then $\bLambda$
can also be recovered from the projected data.

The above arguments are the fundament of the Projected-PCA, and provide
the rationale of our estimators to be defined in Section~\ref{sec2.3}. We shall
make the above arguments rigorous by showing that the projected error
$\bP\bU$ is asymptotically negligible and, therefore, the idiosyncratic
error term $\bU$ can be completely removed by the projection step.

\subsection{Semiparametric factor model}\label{sec2.2}

As one of the useful examples of forming the space $ \bchix$ and the
projection operator, this paper considers
model (\ref{eq1.4}), where $\bX_i$'s and $y_{it}$'s are the only
observable data, and $\{g_k(\cdot)\}_{k\leq K}$ are unknown
nonparametric functions. The specific case (\ref{eq1.2}) (with $\gamma
_{ik}=0$) was used extensively in the financial studies by \citet{CL07},
\citet{CMO} and \citet{park}, with $\bX_i$'s being the observed ``market
characteristic variables.'' We assume $K$ to be known for now. In
Section~\ref{sec6}, we will propose a projected-eigenvalue-ratio method to
consistently estimate $K$ when it is unknown.

We assume that $g_k(\bX_i)$ does not depend on $t$, which means the
loadings represent the cross-sectional heterogeneity only. Such a model
specification is reasonable since in many applications using factor
models, to pertain the stationarity of the time series, the analysis
can be conducted within each fixed time window with either a fixed or
slowly-growing $T$. Through localization in time, it is not stringent
to require the loadings be time-invariant. This also shows one of the
attractive features of our asymptotic results: under mild conditions,
our factor estimates are consistent even if $T$ is finite.


To nonparametrically estimate $g_k(\bX_i)$ without the curse of
dimensionality when $\bX_i$ is multivariate, we assume $g_k(\cdot)$ to
be additive: for each $k \leq K, i\leq p$, there are $(g_{k1},\ldots
,g_{kd})$ nonparametric functions such that
%
\begin{equation}
\label{eq2.3} g_k(\bX_i)=\sum
_{l=1}^d g_{kl}(X_{il}),\qquad  d=\dim(
\bX_i).
\end{equation}
Each additive component of $g_k$ is estimated by the sieve method.
Define $\{\phi_1(x), \phi_2(x),\ldots\}$ to be a set of basis functions
(e.g., B-spline, Fourier series, wavelets, polynomial series), which
spans a dense linear space of the functional space for $\{g_{kl}\}$.
Then for each $l\leq d$,
%
\begin{equation}
\label{eq2.4} g_{kl}(X_{il})=\sum
_{j=1}^J b_{j, kl}\phi_j(X_{il})+
R_{kl}(X_{il}),\qquad  k\leq K, i\leq p, l\leq d.
\end{equation}
Here, $\{b_{j,kl}\}_{j\leq J}$ are the sieve coefficients of the $l$th
additive component of $g_k(\bX_i)$, corresponding to the $k$th factor
loading; $R_{kl}$ is a ``remaining function'' representing the
approximation error; $J$ denotes the number of sieve terms which grows
slowly as $p\rightarrow\infty$. The basic assumption for sieve
approximation is that $\sup_{x}|R_{kl}(x)|\rightarrow0$ as
$J\rightarrow
\infty$. We take the same basis functions in (\ref{eq2.4}) purely for
simplicity of notation.

Define, for each $k \leq K$ and for each $i\leq p$,
\begin{eqnarray*}
 {\mathbf b}_k'&=&( b_{1, k1}, \ldots,
b_{J, k1}, \ldots, b_{1,kd}, \ldots , b_{J,kd}) \in
\mathbb{R}^{Jd},
\\
{\phi}(\bX_i)'&=&\bigl( \phi_1(X_{i1}),
\ldots, \phi_J(X_{i1}), \ldots , \phi _1(X_{id}),
\ldots, \phi_J(X_{id})\bigr) \in\mathbb{R}^{Jd}.
\end{eqnarray*}
Then we can write
\[
g_k(\bX_i)= \phi(\bX_i)'
\mathbf b_k+\sum_{l=1}^d
R_{kl}(X_{il}).
\]
Let $\bB=(\mathbf b_1,\ldots,\mathbf b_{K})$ be a $(Jd)\times K$
matrix of sieve
coefficients, $\Phi(\bX)=(\phi(\bX_1), \ldots, \phi(\bX_p))'$ be a
$p\times(Jd)$ matrix of basis functions, and $\bR(\bX)$ be $p\times K$
matrix with the $(i,k)$th element $\sum_{l=1}^d R_{kl}(X_{il})$. Then
the matrix form of (\ref{eq2.3}) and (\ref{eq2.4}) is
\[
\bG(\bX)=\Phi(\bX)\bB+\bR(\bX).
\]
Substituting this into (\ref{eq1.5}), we write
\[
\bY=\bigl\{\Phi(\bX)\bB+ \bGamma\bigr\} \bF' + \bR(\bX)
\bF' +\bU.
\]
We see that the residual term consists of two parts: the sieve
approximation error $\bR(\bX)\bF'$ and the idiosyncratic $\bU$.
Furthermore, the random effect assumption on the coefficients $\bGamma$
makes it also behave like noise, and hence negligible when the
projection operator $\bP$ is applied.

\subsection{The estimator}\label{sec2.3}

Based on the idea described in Section~\ref{sec2.1}, we propose a Projected-PCA
method, where $\bchix$ is the sieve space spanned by the basis
functions of $\bX$, and $\bP$ is chosen as the projection matrix onto
$\bchix$, defined by the $p\times p$ projection matrix
%
\begin{equation}
\label{eq2.5} \bP=\Phi(\bX) \bigl(\Phi(\bX)'\Phi(\bX)
\bigr)^{-1}\Phi(\bX)'.
\end{equation}
The estimators of the model parameters in (\ref{eq1.5}) are defined as follows.
The columns of $\widehat\bF/\sqrt{T}$ are defined as the eigenvectors
corresponding to the first $K$ largest eigenvalues of the $T\times T$
matrix $\bY'\bP\bY$, and
%
\begin{equation}
\label{eq2.6} \widehat\bG(\bX)=\frac{1}{T}\bP\bY\widehat\bF
\end{equation}
is the estimator of $\bG(\bX)$.

The intuition can be readily seen from the discussions in Section~\ref{sec2.1},
which also provides an alternative formulation of $\widehat\bG(\bX)$ as
follows: let $\widehat\bD$ be a $K\times K$ diagonal matrix consisting
of the largest $K$ eigenvalues of the $p\times p$ matrix $\frac
{1}{T}\bP
\bY\bY'\bP$. Let $\hXi=(\widehat{\bolds{\xi}}_1,\ldots,\widehat{
\bolds{\xi}}_K)$ be a $p\times K$ matrix
whose columns are the corresponding eigenvectors. According to the
relation $(\frac{1}{T}\bP\bY\bY'\bP)\bP\bLambda\approx\bP
\bLambda
(\bLambda'\bP\bLambda)$ described in Section~\ref{sec2.1}, we can also estimate
$\bG(\bX)$ or $\bP\bLambda$ by
\[
\widehat\bG(\bX)=\hXi\widehat\bD^{1/2}.
\]
We shall show in Lemma~\ref{la.1add} that this is equivalent to (\ref{eq2.6}).
Therefore, unlike the traditional PCA method for usual factor models
[e.g., \citet{bai03}, \citet{SW02}], the Projected-PCA takes the
principal components of the projected data $\bP\bY$. The estimator is
thus invariant to the rotation-transformations of the sieve bases. %

The estimation of the loading component $\bGamma$ that cannot be
explained by the covariates can be estimated as follows. With the
estimated factors $\widehat\bF$, the least-squares estimator of
loading matrix is $\widehat{\bLambda} = \bY\widehat{\bF}/T$, by using
(\ref{eq2.1}) and (\ref{eq2.2}). Therefore, by~(\ref{eq1.5}), a natural
estimator of ${\bGamma}$ is
%
\begin{equation}
\label{eq2.7} \widehat{\bGamma} = \widehat{\bLambda} - \widehat{\bG}(\bX) =
\frac
{1}{T}(\bI-\bP)\bY\hF.
\end{equation}
\subsection{Connection with panel data models with time-varying coefficients}\label{sec2.4}

Consider a panel data model with time-varying coefficients as follows:
%
\begin{equation}
\label{eq2.8} y_{it}=\bX_i'
\bbeta_t+\mu_t+u_{it},\qquad  i\leq p, t\leq T,
\end{equation}
where $\bX_i$ is a $d$-dimensional vector of time-invariant regressors
for individual $i$; $\mu_t$ denotes the unobservable random time
effect; $u_{it}$ is the regression error term. The regression
coefficient $\bbeta_t$ is also assumed to be random and time-varying,
but is common across the cross-sectional individuals.

The semiparametric factor model admits (\ref{eq2.8}) as a special case.
Note that (\ref{eq2.8}) can be rewritten as $y_{it}=g(\bX_i)'\bff
_t+u_{it}$ with $K=d+1$ unobservable ``factors'' $\bff_t=(\mu
_t,\bbeta
_t')'$ and ``loading'' $g(\bX_i)=(1,\bX_i')'$. The model (\ref{eq1.4})
being considered, on the other hand, allows more general nonparametric
loading functions.

\section{Projected-PCA in conventional factor models}\label{1541512515}

Let us first consider the asymptotic performance of the Projected-PCA
in the conventional factor model:
%
\begin{equation}
\label{eq3.1} \bY=\bLambda\bF'+\bU.
\end{equation}
In the usual statistical applications for factor analysis, the latent
factors are assumed to be serially independent, while in financial
applications, the factors are often treated to be weakly dependent time
series satisfying strong mixing conditions.

We now demonstrate by a simple example that latent factors $\bF$ can
be estimated at a faster rate of convergence by Projected-PCA than the
conventional PCA and that they can be consistently estimated even when
sample size $T$ is finite.

\begin{exm}\label{ex3.1}
To appreciate the intuition, let us consider a specific case in which
$K = 1$ so that model (\ref{eq1.4}) reduces to
\[
y_{it} = g(X_i) f_t + \gamma_{i}
f_t + u_{it}.
\]
Assume that $g(\cdot)$ is so smooth that it is in fact a constant
$\beta
$ (otherwise, we can use a local constant approximation), where $\beta>
0$. Then the model reduces to
\[
y_{it} = \beta f_t + \gamma_{i}
f_t + u_{it}.
\]
The projection in this case is averaging over $i$, which yields
\[
\bar{y}_{\cdot t} = \beta f_t + \bar{\gamma}_{\cdot}
f_t + \bar {u}_{\cdot t},
\]
where $\bar{y}_{\cdot t}$, $\bar{\gamma}_{\cdot}$ and $\bar
{u}_{\cdot
t}$ denote the averages of their corresponding quantities over $i$. For
the identification purpose, suppose $E\gamma_{i}=Eu_{it}=0$, and $\sum_{t=1}^T f_t^2 = T$. Ignoring the last two terms, we obtain estimators
%
\begin{equation}
\label{e3.2} \hat{\beta} = \Biggl(\frac{1}{T} \sum
_{t=1}^T \bar{y}_{\cdot
t}^2
\Biggr)^{1/2} \quad\mbox{and}\quad \hat{f}_t = \bar{y}_{\cdot t}
/ \hat\beta.
\end{equation}

These estimators are special cases of the Projected-PCA estimators. To
see this, define $\bar\mathbf y=(\bar y_{\cdot1},\ldots, \bar
y_{\cdot T})'$,
and let $\bone_p$ be a $p$-dimensional column vector of ones. Take a
naive basis $\Phi(\bX)=\bone_p$; then the projected data matrix is in
fact $\bP\bY=\bone_p\bar\mathbf y'$. Consider the $T\times T$ matrix
$
\bY'\bP\bY=(\bone_p\bar\mathbf y')'\bone_p\bar\mathbf y'=p\bar
\mathbf y\bar\mathbf y'$,
whose largest eigenvalue is $p\|\bar\mathbf y\|^2$.
From
\[
\bY'\bP\bY\frac{\bar\mathbf y}{\|\bar\mathbf y\|}= p\|\bar \mathbf y\|^2
\frac{\bar\mathbf y
}{\|\bar\mathbf y\|},
\]
we have the first eigenvector of $\bY'\bP\bY$ equals $\bar\mathbf
y/\|\bar\mathbf y
\|$. Hence, the Projected-PCA estimator of factors is $\widehat\bF
=\sqrt
{T}\bar\mathbf y/\|\bar\mathbf y\|$. In addition, the Projected-PCA
estimator of
the loading vector $\beta\bone_p$ is
\[
\frac{1}{T}\bone_p\bar\mathbf y'\widehat\bF=
\frac{1}{\sqrt
{T}}\bone_p\|\bar\mathbf y \|.
\]
Hence, the Projected-PCA-estimator of $\beta$ equals $\|\bar\mathbf
y\|/\sqrt
{T}$. These estimators match with (\ref{e3.2}). Moreover, since the
ignored two terms $\bar\gamma_{\cdot}$ and $\bar u_{\cdot t}$ are of
order $O_p(p^{-1/2})$, $\hat{\beta}$ and $\hat{f}_t$ converge whether
or not $T$ is large. Note that this simple example satisfies all the
assumptions to be stated below, and $\widehat\beta$ and $\widehat f_t$
achieve the same rate of convergence as that of Theorem~\ref{th4.1}. We
shall present more details about this example in Appendix G in the
supplementary material [\citet{PPCAsupp}]. 
\end{exm}

\subsection{Asymptotic properties of Projected-PCA}\label{1541512515.1}

We now state the conditions and results formally in the more general
factor model (\ref{eq3.1}).
Recall that the projection matrix is defined as
\[
\bP=\Phi(\bX) \bigl(\Phi(\bX)'\Phi(\bX)\bigr)^{-1}\Phi(
\bX)'.
\]

The following assumption is the key condition of the Projected-PCA.

\begin{assum}[(Genuine projection)]\label{ass3.1}
There are positive constants $c_{\min}$ and $c_{\max}$ such that, with
probability approaching one (as $p\to\infty$),
\[
c_{\min}<\lambda_{\min}\bigl(p^{-1}
\bLambda'\bP\bLambda\bigr)<\lambda _{\max
}
\bigl(p^{-1}\bLambda'\bP\bLambda\bigr)<c_{\max}.
\]
\end{assum}

Since the dimensions of $\Phi(\bX)$ and $\bLambda$ are, respectively,
$p\times Jd$ and $p\times K$, Assumption~\ref{ass3.1} requires $Jd\geq
K$, which is reasonable since we assume $K$, the number of factors, to
be fixed throughout the paper.

Assumption~\ref{ass3.1} is similar to the \textit{pervasive} condition
on the factor loadings [\citet{SW02}]. In our context, this condition
requires the covariates $\bX$ have nonvanishing explaining power on the
loading matrix, so that the projection matrix $\bLambda'\bP\bLambda$
has spiked eigenvalues. Note that it rules out the case when $\bX$ is
completely unassociated with the loading matrix $\bLambda$ (e.g., when
$\bX$ is pure noise). One of the typical examples that satisfies this
assumption is the semiparametric factor model [model (\ref{eq1.4})]. We
shall study this specific type of factor model in Section~\ref{s4}, and prove
Assumption~\ref{ass3.1} in the supplementary material [\citet{PPCAsupp}].

Note that $\bF$ and $\bLambda$ are not separately identified, because
for any nonsingular $\bH$, $\bLambda\bF'=\bLambda\bH^{-1}\bH\bF'$.
Therefore, we assume the following.

\begin{assum}[(Identification)]\label{ass3.2}
Almost surely,
$T^{-1}\bF'\bF=\bI_K$ and $\bLambda'\bP\bLambda$ is a $K\times K$
diagonal matrix with distinct entries.
\end{assum}


This condition corresponds to the PC1 condition of \citet{BN13}, which
separately identifies the factors and loadings from their product
$\bLambda\bF'$.
It is often used in factor analysis for identification, and means that
the columns of factors and loadings can be orthogonalized [also see
\citet{Bli12}]. 

\begin{assum}[(Basis functions)]\label{ass3.3}
(i) There are $d_{\min}$ and $d_{\max}>0$ so that with probability
approaching one (as $p\to\infty$),
\[
d_{\min}<\lambda_{\min}\bigl(p^{-1}\Phi(
\bX)'\Phi(\bX)\bigr)<\lambda _{\max
}\bigl(p^{-1}
\Phi(\bX)'\Phi(\bX)\bigr)<d_{\max}.
\]

(ii) $\max_{j\leq J, i\leq p, l\leq d}E\phi_j(X_{il})^2<\infty$.
\end{assum}

Note that $p^{-1}\Phi(\bX)'\Phi(\bX)=p^{-1}\sum_{i=1}^p\phi(\bX
_i)'\phi
(\bX_i)$ and $\phi(\bX_i)$ is a vector of dimensionality $J d \ll p$.
Thus, condition (i) can follow from the strong law of large numbers.
For instance, $\{\bX_i\}_{i\leq p}$ are weakly correlated and in the
population level $E\phi(\bX_i)'\phi(\bX_i)$ is well-conditioned. In
addition, this condition can be satisfied through proper normalizations
of commonly used basis functions such as B-splines, wavelets, Fourier
basis, etc. In the general setup of this paper, we allow $\{\bX_i\}
_{i\leq p}$'s to be cross-sectionally dependent and nonstationary.
Regularity conditions about weak dependence and stationarity are
imposed only on $\{(\bff_t, \mathbf u_t)\}$ as follows.

We impose the strong mixing condition. Let $\mathcal{F}_{-\infty}^0$
and $\mathcal{F}_{T}^{\infty}$ denote the $\sigma$-algebras generated
by $\{(\bff_t,\mathbf u_t): t\leq0\}$ and $\{(\bff_t,\mathbf u_t):
t\geq T\}$,
respectively. Define the mixing coefficient
\[
\alpha(T)=\sup_{A\in\mathcal{F}_{-\infty}^0, B\in\mathcal
{F}_{T}^{\infty
}}\bigl|P(A)P(B)-P(AB)\bigr|.
\]

\begin{assum}[(Data generating process)] \label{ass3.4}
(i) $\{\mathbf u_t, \bff_t\}_{t\leq T}$ is strictly stationary.
In addition, $Eu_{it}=0$ for all $i\leq p, j\leq K$; $\{\mathbf u_t\}
_{t\leq
T}$ is independent of $\{\bX_i, \bff_t\}_{i\leq p, t\leq T}$.

\begin{longlist}[(iii)]
\item[(ii)] Strong mixing: there exist $r_1, C_1>0$ such that for all $T>0$,
\[
\alpha(T)<\exp\bigl(-C_1T^{r_1}\bigr).
\]

\item[(iii)]
Weak dependence: there is $C_2>0$ so that
\begin{eqnarray*}
\max_{j\leq p}\sum_{i=1}^p|Eu_{it}u_{jt}|&<&C_2,\\
\frac{1}{pT}\sum_{i=1}^p\sum
_{j=1}^p\sum_{t=1}^T
\sum_{s=1}^T|Eu_{it}u_{js}|&<&C_2,
\\
\max_{i\leq p}\frac{1}{pT}\sum_{k=1}^p
\sum_{m=1}^p \sum
_{t=1}^T\sum_{s=1}^T\bigl|
\cov(u_{it}u_{kt}, u_{is}u_{ms})\bigr|&<& C_2.
\end{eqnarray*}
%
\item[(iv)] Exponential tail: there exist $r_2, r_3>0$ satisfying
$r_1^{-1}+r_2^{-1}+r_3^{-1}>1$ and $b_1, b_2>0$, such that for any
$s>0$, $i\leq p$ and $j\leq K$,
\[
P\bigl(|u_{it}|>s\bigr)\leq\exp\bigl(-(s/b_1)^{r_2}\bigr),\qquad
P\bigl(|f_{jt}|>s\bigr)\leq\exp \bigl(-(s/b_2)^{r_3}\bigr).
\]
\end{longlist}
\end{assum}

Assumption~\ref{ass3.4} is standard, especially condition (iii) is
commonly imposed for high-dimensional factor analysis [e.g., \citet
{SW02,bai03}], which requires $\{u_{it}\}_{i\leq p, t\leq T}$ be weakly
dependent
both serially and cross-sectionally. It is often satisfied when the
covariance matrix $E\mathbf u_t\mathbf u_t'$ is sufficiently sparse
under the
strong mixing condition. We provide primitive conditions of condition
(iii) in the supplementary material [\citet{PPCAsupp}].

Formally, we have the following theorem:

\begin{thmm}\label{th3.1}
Consider the conventional factor model (\ref{eq3.1}) with
Assumptions \ref{ass3.1}--\ref{ass3.4}. The Projected-PCA estimators
$\hF$ and $\hG(\bX)$ defined in Section~\ref{sec2.3} satisfy, as $p
\rightarrow\infty$ [$J, T$ may either grow simultaneously with $p$
satisfying $J=o(\sqrt{p})$ or stay constant with $Jd\geq K$],
%
\begin{eqnarray}
 \frac{1}{T}\|\hF-\bF\|_F^2&=&
O_P \biggl(\frac{J}{p} \biggr),
\\
\frac{1}{p}\bigl\|\hG(\bX)-\bP\bLambda\bigr\|_F^2&=&
O_P \biggl(\frac
{J}{pT}+\frac
{J^2}{p^2} \biggr).\label{eq3.4add}
\end{eqnarray}
%
\end{thmm}

To compare with the traditional PCA method, the convergence rate for
the estimated factors is improved for small $T$. In particular, the
Projected-PCA does not require $T\rightarrow\infty$, and also has a
good rate of convergence for the loading matrix up to a projection
transformation. Hence, we have achieved a finite-$T$ consistency, which
is particularly interesting in the ``high-dimensional-low-sample-size''
(HDLSS) context, considered by \citet{JM09}. In contrast, the
traditional PCA method achieves a rate of convergence of
$O_P(1/p+1/T^2)$ for estimating factors, and $O_P(1/T+1/p)$ for
estimating loadings.
See Remarks \ref{re4.1}, \ref{re4.2} below for additional details.

Let $\bSigma=\cov(\mathbf y_t)$ be the $p\times p$ covariance matrix
of $\mathbf y
_t=(y_{1t},\ldots,y_{pt})'$. Convergence (\ref{eq3.4add}) in Theorem~\ref
{th3.1} also describes the relationship between the leading
eigenvectors of $\frac{1}{T}\bP\bY\bY'\bP$ and those of $\bSigma
$. To
see this, let $\bXi=(\bolds{\xi}_1,\ldots,\bolds{\xi}_K)$ be the
eigenvectors of $\bSigma
$ corresponding to the first $K$ eigenvalues. Under the \textit
{pervasiveness condition}, $\bXi$ can be approximated by $\bLambda$
multiplied by a positive definite matrix of transformation [\citet
{POET}]. In the context of Projected-PCA, by definition, $\widehat\bXi
=\widehat\bG(\bX)\widehat\bD^{-1/2}$; here we recall that
$\widehat\bD
$ is a diagonal matrix consisting of the largest $K$ eigenvalues of
$\frac{1}{T}\bP\bY\bY'\bP$, and $\hXi$ is a $p\times K$ matrix whose
columns are the corresponding eigenvectors. Then (\ref{eq3.4add})
immediately implies the following corollary, which complements the PCA
consistency in \textit{spiked covariance models} [e.g., \citet
{johnstone2001distribution} and \citet{paul2007asymptotics}].

\begin{cor}\label{th3.2} Under the conditions of Theorem~\ref{th3.1},
there is a $K\times K$ positive definite matrix $\bV$, whose
eigenvalues are bounded away from both zero and infinity, so that as
$p\to\infty$ [$J, T$ may either grow simultaneously with $p$ satisfying
$J=o(\sqrt{p})$ or stay constant with $Jd\geq K$],
\[
\|\widehat\bXi-\bXi\bV\|_F= O_P \biggl(
\frac{1}{p}\|\bSigma_u\|_2+ \sqrt {
\frac{J}{pT}}+\frac{J}{p}+\frac{1}{\sqrt{p}}\|\bP\bLambda -\bLambda\|
_F \biggr).
\]
\end{cor}

\section{Projected-PCA in semiparametric factor models}\label{s4}

\subsection{Sieve approximations}\label{sec4.1}

In the semiparametric factor model, it is assumed that $\lambda
_{ik}=g_k(\bX_i)+\gamma_{ik}$, where $g_k(\bX_i)$ is a nonparametric
smooth function for the observed covariates, and $\gamma_{ik}$ is the
unobserved random loading component that is independent of $\bX_i$.
Hence, the model is written as
\[
y_{it}=\sum_{k=1}^K \bigl
\{g_k(\bX_{i})+\gamma_{ik} \bigr\}
f_{tk}+u_{it},\qquad i=1,\ldots,p, t=1,\ldots,T.
\]
In the matrix form,
\[
\bY=\bigl\{\bG(\bX)+ \bGamma\bigr\}\bF' +\bU,
\]
and $\bG(\bX)$ does not vanish (pervasive condition; see Assumption~\ref
{ass4.2} below). 

The estimators $\hF$ and $\hG(\bX)$ are the Projected-PCA estimators as
defined in Section~\ref{sec2.3}. We now define the estimator of the
nonparametric function $g_k(\cdot)$, $k=1,\ldots,K$.
In the matrix form, the projected data has the following sieve
approximated representation:
%
\begin{equation}
\label{eq4.1} \bP\bY=\Phi(\bX)\bB\bF'+ \widetilde\bE,
\end{equation}
where $ \widetilde\bE=\bP\bGamma\bF' +\bP\bR(\bX)\bF' +\bP\bU
$ is
``small'' because $\bGamma$ and $\bU$ are orthogonal to the function
space spanned by $\bX$, and $\bR(\bX)$ is the sieve approximation error.
The sieve coefficient matrix $\bB=(\mathbf b_1,\ldots,\mathbf b_K)$
can be estimated
by least squares from the projected model (\ref{eq4.1}): Ignore
$\widetilde\bE$, replace $\bF$ with $\hF$, and solve (\ref{eq4.1})
to obtain
\[
\widehat\bB=(\hb_1,\ldots,\hb_K)=\frac{1}{T}\bigl[
\Phi(\bX)'\Phi (\bX)\bigr]^{-1}\Phi (\bX)'
\bY\hF.
\]
We then estimate $g_k(\cdot)$ by
\[
\widehat g_k(\mathbf x)=\phi(\mathbf x)'
\hb_k \qquad\forall\mathbf x\in\mathcal{X}, k=1,\ldots,K,
\]
where $\mathcal{X}$ denotes the support of $\bX_i$. 

\subsection{Asymptotic analysis}\label{sec4.2}

When $\bLambda=\bG(\bX)+\bGamma$, $\bG(\bX)$ can be understood as the
projection of $\bLambda$ onto the sieve space spanned by $\bX$. Hence,
the following assumption is a specific version of Assumptions \ref
{ass3.1} and \ref{ass3.2} in the current context.


\begin{assum} \label{ass4.1}
(i) Almost surely, $T^{-1}\bF'\bF=\bI_K$ and $\bG(\bX)'\bG(\bX)$
is a
$K\times K$ diagonal matrix with distinct entries.

(ii) There are two positive constants $c_{\min}$ and $c_{\max}$ so that
with probability approaching one (as $p\to\infty$),
\[
c_{\min}<\lambda_{\min}\bigl(p^{-1}\bG(
\bX)'\bG(\bX)\bigr)<\lambda_{\max
}\bigl(p^{-1}\bG(
\bX)'\bG(\bX)\bigr)<c_{\max}.
\]
\end{assum}

In this section, we do not need to assume $\{\bolds{\gamma}_i\}
_{i\leq p}$ to
be i.i.d. for the estimation purpose. Cross-sectional weak dependence
as in Assumption~\ref{ass4.2}(ii) below would be sufficient. The i.i.d.
assumption will be only needed when we consider specification tests in
Section~\ref{sec5}. Write $\bolds{\gamma}_i=(\gamma_{i1},\ldots,\gamma
_{iK})'$, and
\[
\nu_p=\max_{k\leq K} \frac{1}p \sum
_{i\le p}\var(\gamma_{ik}).
\]

\begin{assum} \label{ass4.2}
(i) $E\gamma_{ik}=0$ and $\{\bX_i\}_{i\leq p}$ is independent of $\{
\gamma_{ik}\}_{i\leq p}$.

(ii) $\max_{k\leq K, i\leq p}Eg_k(\bX
_i)^2<\infty$, $\nu_p<\infty$ and
\[
\max_{k \le K, j\leq p}\sum_{i\leq p}|E
\gamma_{ik}\gamma_{jk} |=O(\nu_p).
\]
%
\end{assum}


The following set of conditions is concerned about the accuracy of the
sieve approximation.

\begin{assum}[(Accuracy of sieve approximation)]\label{ass4.3} $\forall
l\leq d, k\leq K$,

(i) the loading component $g_{kl}(\cdot)$ belongs to a H\"{o}lder
class $\mathcal{G}$ defined by
\[
\mathcal{G}=\bigl\{g:\bigl |g^{(r)}(s)-g^{(r)}(t)\bigr|\leq
L|s-t|^{\alpha} \bigr\}
\]
for some $L > 0$;

(ii) the sieve coefficients $\{b_{k,jl}\}_{j\leq J}$ satisfy for
$\kappa=2(r+\alpha)\geq4$, as $J\rightarrow\infty$,
\[
\sup_{x\in\mathcal{X}_l}\Biggl|g_{kl}(x)-\sum
_{j=1}^Jb_{k,jl}\phi _j(x)\Biggr|^2=O
\bigl(J^{-\kappa}\bigr),
\]
where $\mathcal{X}_l$ is the support of the $l$th element of $\bX_i$,
and $J$ is the sieve dimension.

(iii) $\max_{k, j, l} b_{k, jl}^2 <\infty$.
\end{assum}

Condition (ii) is satisfied by common basis. For example, when $\{\phi
_j\}$ is polynomial basis or B-splines, condition (ii) is implied by
condition (i) [see, e.g., \citet{lorentz} and \citet{chen}].

\begin{thmm}\label{th4.1}
Suppose $J=o(\sqrt{p})$. Under Assumptions \ref{ass3.3}, \ref{ass3.4},
\ref{ass4.1}--\ref{ass4.3}, as $p, J\rightarrow\infty$,
$T$ can be either divergent or bounded, we have that
\begin{eqnarray*}
\frac{1}{T}\|\hF-\bF\|_F^2&=&O_P
\biggl(\frac{1}{p}+\frac
{1}{J^{\kappa}} \biggr) ,
\\
\frac{1}{p}\bigl\|
\widehat\bG(\bX)-\bG(\bX)\bigr\|_F^2&=& O_P \biggl(
\frac{J}{p^2} +\frac{J}{pT}+\frac{J}{J^{\kappa}}+\frac{J\nu_p}{p}
\biggr),
\\
\max_{k\leq K}\sup_{\mathbf x\in\mathcal{X}}\bigl|\widehat
g_k(\mathbf x)-g_k(\mathbf x)\bigr| &=&O_P
\biggl(\frac{J}{p}+ \frac{J}{\sqrt{pT}}+\frac{J}{J^{\kappa
/2}}+J\sqrt {
\frac{\nu_p}{p}} \biggr) \max_{j\leq J}\sup_x\bigl|
\phi_j(x)\bigr|.
\end{eqnarray*}
In addition, if $T\rightarrow\infty$ simultaneously with $p$ and $J$, then
\[
\frac{1}{p}\|\widehat\bGamma-\bGamma\|_F^2=O_P
\biggl(\frac{J}{p^2} + \frac
{1}{T} + \frac{1}{J^{\kappa}}+
\frac{J\nu_p}{p} \biggr).
\]
\end{thmm}

The optimal $J^*=(p\min\{T,p,\nu_p^{-1}\})^{1/\kappa}$ simultaneously
minimizes the convergence rates of the factors and nonparametric
loading function $g_k(\cdot)$. It also satisfies the constraint
$J^*=o(\sqrt{p})$ as $\kappa\geq4$. With $J=J^*$, we have
\begin{eqnarray*}
\frac{1}{T}\sum_{t=1}^T\|\widehat
\bff_t-\bff_t\|^2&=&O_P \biggl(
\frac{1}{p} \biggr) ,
\\
\frac{1}{p}\sum
_{i=1}^p\bigl|\widehat g_k(
\bX_i)-g_k(\bX_i)\bigr|^2&=&
O_P \biggl(\frac{1}{(p\min\{T, p, v_p^{-1}\})^{1-1/\kappa}} \biggr)
\qquad\forall k,
\\
\max
_{k\leq K}\sup_{\mathbf x\in\mathcal{X}}\bigl|\widehat g_k(
\mathbf x)- g_k(\mathbf x)\bigr| &=& O_P \biggl(
\frac{ \max_{j\leq J}\sup_x|\phi_j(x)| }{(p\min\{T, p,
\nu
_p^{-1}\})^{1/2-1/\kappa}} \biggr),
\end{eqnarray*}
and $\widehat\bGamma=(\widehat{\bolds{\gamma}}_1,\ldots,\widehat
{\bolds{\gamma}}_p)'$ satisfies
\[
\frac{1}{p}\sum_{i=1}^p\|
\widehat{\bolds{\gamma}}_i-\bolds {\gamma}_i
\|^2= O_P \biggl(\frac{1}{(p\min\{T, p, v_p^{-1}\})^{1-1/\kappa}}+\frac{1}{T}
\biggr).
\]

Some remarks about these rates of convergence compared with those of
the conventional factor analysis are in order.

\begin{remark}\label{re4.1}
The rates of convergence for factors and nonparametric functions do not
require $T\rightarrow\infty$. When $T=O(1)$,
\[
\frac{1}{T}\sum_{t=1}^T\|\widehat
\bff_t-\bff_t\|^2=O_P \biggl(
\frac
{1}{p} \biggr), \qquad\frac{1}{p}\sum_{i=1}^p\bigl|
\widehat g_k(\bX_i)-g_k(
\bX_i)\bigr|^2= O_P \biggl(\frac{1}{p^{1-1/\kappa}}
\biggr).
\]
The rates still converge fast when $p$ is large, demonstrating the
blessing of dimensionality. This is an attractive feature of the
Projected-PCA in the HDLSS context, as in many applications, the
stationarity of a time series and the time-invariance assumption on the
loadings hold only for a short period of time. In contrast, in the
usual factor analysis, consistency is granted only when $T\rightarrow
\infty$. For example, according to \citet{FLS14} (Lemma C.1), the
regular PCA method has the following convergence rate:
\[
\frac{1}{T}\sum_{t=1}^T\|\widetilde
\bff_t-\bff_t\|^2=O_P \biggl(
\frac
{1}{p}+\frac{1}{T^2} \biggr),
\]
which is inconsistent when $T$ is bounded.
\end{remark}

\begin{remark}\label{re4.2}
When both $p$ and $T$ are large, the Projected-PCA estimates factors
as well as the regular PCA does, and achieves a faster rate of
convergence for the estimated loadings when $\gamma_{ik}$ vanishes. In
this case, $\lambda_{ik} = g_k(\bX_i)$, the loading matrix is estimated
by $\widehat{\bLambda}=\widehat\bG(\bX) $, and
\[
\frac{1}{p}\sum_{i=1}^p|\widehat
\lambda_{ik}- \lambda_{ik}|^2 = \frac
{1}{p}
\sum_{i=1}^p\bigl|\widehat g_k(
\bX_i)-g_k(\bX_i)\bigr|^2=
O_P \biggl(\frac
{1}{(pT)^{1-1/\kappa}}+\frac{1}{p^{2-2/\kappa}} \biggr).
\]
In contrast, the regular PCA method as in \citet{SW02} yields
\[
\frac{1}{p}\sum_{i=1}^p|\widetilde
\lambda_{ik}- \lambda_{ik}|^2= O_P
\biggl( \frac{1}{T} +\frac{1}{p} \biggr).
\]
Comparing these rates, we see that when $g_k(\cdot)$'s are sufficiently
smooth (larger $\kappa$), the rate of convergence for the estimated
loadings is also improved.
\end{remark}

\section{Semiparametric specification test}\label{sec5}


The loading matrix always has the following orthogonal decomposition:
\[
\bLambda=\bG(\bX)+\bGamma,
\]
where $\bGamma$ is interpreted as the loading component that cannot be
explained by $\bX$.
We consider two types of specification tests: testing $H_0^1: \bG(\bX
)=0$, and $H_0^2: \bGamma=0$. The former tests whether the observed
covariates have explaining powers on the loadings, while the latter
tests whether the covariates fully explain the loadings. The former
provides a diagnostic tool as to whether or not to employ the
Projected-PCA; the latter tests the adequacy of the semiparametric
factor models in the literature.

\subsection{Testing $\bG(\bX)=0$}\label{sec5.1}
Testing whether the observed covariates have explaining powers on the
factor loadings can be formulated as the following null hypothesis:
\[
H_0^1: \bG(\bX)=0\qquad \mbox{a.s.}
\]
Due to the approximate orthogonality of $\bX$ and $\bGamma$, we have
$\bP\bLambda\approx\bG(\bX)$. Hence, 
the null hypothesis is approximately equivalent to
\[
H_0:\bP\bLambda=0 \qquad\mbox{a.s.}
\]
This motivates a statistic $\|\bP\widetilde\bLambda\|_F^2 = \tr(
\widetilde\bLambda' \bP\widetilde\bLambda)$ for a consistent loading
estimator $\widetilde\bLambda$. Normalizing the test statistic by its
asymptotic variance leads to the test statistic
\[
S_G=\frac{1}{p}\tr\bigl(\bW_1\widetilde
\bLambda'\bP\widetilde\bLambda \bigr),\qquad \bW_1=\biggl(
\frac{1}{p}\widetilde\bLambda'\widetilde\bLambda
\biggr)^{-1},
\]
where the $K\times K$ matrix $\bW_1$ is the weight matrix.
The null hypothesis is rejected when $S_G$ is large.

The Projected-PCA estimator is inappropriate under the null hypothesis
as the projection is not genuine. We therefore use the least squares
estimator $\widetilde\bLambda=\bY\widetilde\bF/T$, leading to the
test statistic
%
\[
S_G=\frac{1}{T^2p}\tr\bigl(\bW_1\widetilde
\bF'\bY'\bP\bY\widetilde \bF\bigr).
\]
%
Here, we take $\widetilde\bF$ as the traditional PCA estimator: the
columns of $\widetilde\bF/\sqrt{T}$ are the first $K$ eigenvectors of
the $T\times T$ data matrix $\bY'\bY$.



\subsection{Testing \texorpdfstring{$\bGamma=0$}{$Gamma=0$}}\label{sec5.2}

Connor, Hagmann and Linton (\citeyear{CMO}) applied the semiparametric factor model to analyzing
financial returns, who assumed that
$\bGamma=0$, that is, the loading matrix can be fully explained by the
observed covariates.
It is therefore natural to test the following null hypothesis of
specification:
\[
H_0^2: \bGamma=0 \qquad\mbox{a.s.}
\]
Recall that $ \bG(\bX) \approx\bP\bLambda$ so that $\bLambda
\approx
\bP\bLambda+\bGamma$.
Therefore, essentially the specification testing problem is equivalent
to testing
\[
H_0: \bP\bLambda=\bLambda\qquad  \mbox{a.s.}
\]
That is, we are testing whether the loading matrix in the factor model
belongs to the space spanned by the observed covariates.

A natural test statistic is thus based on the weighted quadratic form
\[
\tr\bigl(\widehat\bGamma'\bW_2\widehat\bGamma\bigr)=
\tr\bigl(\widehat\bLambda '(\bI -\bP)'
\bW_2 (\bI-\bP) \widehat\bLambda\bigr),
\]
for some $p \times p$ positive definite weight matrix $\bW_2$, where
$\hF$ is the Projected-PCA estimator for factors and $\widehat
\bLambda
=\bY\hF/T$.
To control the size of the test, we take $\bW_2=\Sig_u^{-1}$, where
$\Sig_u$ is a diagonal covariance matrix of $\mathbf u_t$ under $H_0$,
assuming that $(u_{1t},\ldots,u_{pt})$ are uncorrelated.

We replace $\Sig_u^{-1}$ with its consistent estimator: let $\widehat
\bU
=\bY-\widehat\bLambda\hF'$. Define
\[
\widehat\Sig_u= T^{-1} \diag\bigl\{\widehat\bU\widehat
\bU'\bigr\}= T^{-1} \diag\bigl\{ \bY\bigl(
\bI-T^{-1}\hF\hF'\bigr)\bY'\bigr\}.
\]
Then the operational test statistic is defined to be
\[
S_{\Gamma}= \tr\bigl(\widehat\bLambda'(\bI-
\bP)'\widehat\Sig _u^{-1}(\bI-\bP ) \widehat
\bLambda\bigr).
\]
The null hypothesis is rejected for large values of $S_{\Gamma}$.

\subsection{Asymptotic null distributions}\label{sec5.3}



For the testing purpose, we assume $\{\bX_i, \bolds{\gamma}_i\}$ to
be i.i.d.,
and let $T, p, J\rightarrow\infty$ simultaneously.
The following assumption regulates the relation between $T$ and $p$.

\begin{assum}\label{ass5.1}
Suppose
(i) $\{\bX_i, \bolds{\gamma}_i\}_{i\leq p}$ are independent and identically
distributed;\vspace*{-6pt}

\begin{longlist}[(iii)]
\item[(ii)] $T^{2/3}=o(p)$, and $p(\log p)^4=o(T^2)$;

\item[(iii)] $J$ and $\kappa$ satisfy: $J=o(\min\{\sqrt{p}, \sqrt{T} \})$, and
$\max\{T\sqrt{p},p\} = o(J^\kappa)$.
\end{longlist}
\end{assum}

Condition (ii) requires a balance of the dimensionality and the sample
size. On one hand, a relatively large sample size is desired [$p(\log
p)^4=o(T^2)$] so that the effect of estimating $\Sig_u^{-1}$ is
negligible asymptotically. On the other hand, as is common in
high-dimensional factor analysis, a lower bound of the dimensionality
is also required [condition $T^{2/3}=o(p)$] to ensure that the factors
are estimated accurately enough. Such a required balance is common for
high-dimensional factor analysis [e.g., \citet{bai03}, \citet{SW02}] and
in the recent literature for PCA [e.g., \citet{JM09}, \citet{SSZM}]. {The
i.i.d. assumption of covariates $\bX_i$ in condition (i) can be relaxed
with further distributional assumptions on $\bolds{\gamma}_i$ (e.g., assuming
$\bolds{\gamma}_i$ to be Gaussian). The conditions on $J$ in
condition (iii)
is consistent with those of the previous sections.}

We focus on the case when $\mathbf u_{t}$ is Gaussian, and show that
under $H_0^1$,
\[
S_G=\bigl(1+o_P(1)\bigr)\frac{1}{p}\tr\bigl(
\bW_1\bGamma'\bP\bGamma\bigr),
\]
and under $H_0^2$
\[
S_{\Gamma}=\bigl(1+o_P(1)\bigr)\frac{1}{T^2}\tr\bigl(
\bF'\bU'\Sig_u^{-1}\bU\bF
\bigr),
\]
whose conditional distributions (given $\bF$) under the null are $\chi
^2$ with degree of freedom, respectively, $JdK$ and $pK$. We can derive
their standardized limiting distribution as $J, T, p\rightarrow\infty$.
This is given in the following result.

\begin{thmm}\label{th5.1}
Suppose Assumptions \ref{ass3.3}, \ref{ass3.4}, \ref{ass4.2}, \ref
{ass5.1} hold. Then under $H_0^1$,
\[
\frac{pS_{G} -JdK}{\sqrt{2JdK}}\mathop{\rightarrow}^d N(0,1),
\]
where $K=\dim(\bff_t)$ and $d=\dim(\bX_i)$.
In addition, suppose Assumptions \ref{ass4.1} and~\ref{ass4.3} further
hold, $\{\mathbf u_t\}_{t\leq T}$ is i.i.d. $N(0,\Sig_u)$ with a diagonal
covariance matrix $\Sig_u$ whose elements are bounded away from zero
and infinity. 
Then under $H_0^2$,
\[
\frac{TS_{\Gamma}-pK}{\sqrt{2pK}}\mathop{\rightarrow}^d N(0,1).
\]
\end{thmm}

In practice, when a relatively small sieve dimension $J$ is used, one
can instead use the upper $\alpha$-quantile of the $\chi^2_{JdK}$
distribution for $pS_G$. 

\begin{remark}
We require $u_{it}$ be independent across $t$, which ensures that the
covariance matrix of the leading term $\vecc(\frac{1}{\sqrt{T}}\bU
\bF
')$ to have a simple form $\Sig_u^{-1}\otimes\bI_K$. This assumption
can be relaxed to allow for weakly dependent $\{\mathbf u_t\}_{t\leq
T}$, but
many autocovariance terms will be involved in the covariance matrix.
One may regularize standard autocovariance matrix estimators such as
\citet{NW87} and \citet{andrews} to account for the high dimensionality.
Moreover, we assume $\Sig_u$ be diagonal to facilitate estimating
$\Sig
_u^{-1}$, which can also be weakened to allow for a nondiagonal but
sparse $\Sig_u$. 
Regularization methods such as thresholding [\citet{Bickel08a}] can then
be employed, though they are expected to be more technically involved.
\end{remark}


\section{Estimating the number of factors from projected data}\label{sec6}

We now address the problem of estimating $K=\dim(\bff_t)$ when it is
unknown. Once a consistent estimator of $K$ is obtained, all the
results achieved carry over to the unknown $K$ case using a
conditioning argument.\setcounter{footnote}{1}\footnote{One can first conduct the analysis
conditioning on the event $\{\widehat K=K\}$, then argue that the
results still hold unconditionally as $P(\widehat K=K)\to1$.} In
principle, many consistent estimators of $K$ can be employed, for
example, \citet{BN02}, \citet{ABC}, \citet{BP09}, \citet{HL}. More
recently, \citet{AH} and \citet{LamYao} proposed to select the largest
ratio of the adjacent eigenvalues of $\bY'\bY$, based on the fact that
the $K$ largest eigenvalues of the sample covariance matrix grow as
fast as $p$ increases, while the remaining eigenvalues either remain
bounded or grow slowly. 

We extend Ahn and Horenstein's (\citeyear{AH}) theory in two ways. First, when the loadings
depend on the observable characteristics, it is more desirable to work
on the projected data $\bP\bY$.
Due to the orthogonality condition of $\bU$ and $\bX$, the projected
data matrix is approximately equal to $\bG(\bX)\bF'$. The projected
matrix $\bP\bY(\bP\bY)'$ thus allows us to study the eigenvalues of
the principal matrix component $\bG(\bX)\bG(\bX)'$, which directly
connects with the strengths of those factors. Since the nonvanishing
eigenvalues of $\bP\bY(\bP\bY)'$ and $(\bP\bY)'\bP\bY= \bY'
\bP\bY$
are the same,
we can work directly with the eigenvalues of the matrix $\bY' \bP\bY
$. Second, we allow $p/T\rightarrow\infty$. 

Let $\lambda_k(\bY'\bP\bY)$ denote the $k$th largest eigenvalue of the
projected data matrix $\bY'\bP\bY$. We assume $0 < K < Jd/2$, which
naturally holds if the sieve dimension $J$ slowly grows. 
The estimator is defined as
\[
\widehat K=\arg\max_{0< k< Jd/2}\frac{\lambda_k(\bY' \bP\bY
)}{\lambda
_{k+1}(\bY' \bP\bY)}.
\]
%

The following assumption is similar to that of \citet{AH}. Recall that
$\bU=(\mathbf u_1,\ldots,\mathbf u_T)$ is a $p\times T$ matrix of the
idiosyncratic
components, and $\Sig_u=E\mathbf u_t\mathbf u_t'$ denotes the $p\times
p $
covariance matrix of $\mathbf u_t$.

\begin{assum}\label{ass6.1}
The error matrix $\bU$ can be decomposed as
%
\begin{equation}
\label{eq5.1}\bU= {\bolds\Sigma}_u^{1/2} \bE
\bM^{1/2},
\end{equation}
where:
\begin{longlist}[(iii)]
\item[(i)] the eigenvalues of $\Sig_u$ are bounded away from zero and
infinity,

\item[(ii)] $\bM$ is a $T$ by $T$ positive semidefinite nonstochastic matrix,
whose eigenvalues are bounded away from zero and infinity,

\item[(iii)] $\bE=(e_{it})_{p\times T}$ is a $p\times T$ stochastic matrix,
where $e_{it}$ is independent in both $i$ and $t$, and $\mathbf e_t =
(e_{1t}, \ldots, e_{pt})'$ are i.i.d. isotropic sub-Gaussian vectors,
that is, there is $C>0$, for all $s>0$,
\[
\sup_{\|\mathbf v\|=1} P\bigl(\bigl|\mathbf v'\mathbf
e_t\bigr|>s\bigr)\leq\exp\bigl(1-Cs^2\bigr).
\]
\item[(iv)] There are $d_{\min}, d_{\max}>0$, almost surely,
\[
d_{\min} \leq\lambda_{\min}\bigl(\Phi(\bX)'\Phi(
\bX)/p\bigr)\leq\lambda _{\max
}\bigl(\Phi(\bX)'\Phi(\bX)/p
\bigr) \le d_{\max}.
\]
\end{longlist}
\end{assum}

This assumption allows the matrix $\bU$ to be both cross-sectionally
and serially dependent. The $T\times T$ matrix $\bM$ captures the
serial dependence across $t$. In the special case of
no-serial-dependence, the decomposition (\ref{eq5.1}) is satisfied by taking
$\bM=\bI$.
In addition, we require $\mathbf u_t$ to be sub-Gaussian to apply random
matrix theories of \citet{vershynin2010introduction}. For instance, when
$\mathbf u_t$ is $N(\bzero,\Sig_u)$, for any $\|\mathbf v\|=1$,
$\mathbf v'\mathbf e_t\sim
N(0,1)$, and thus condition (iii) is satisfied. Finally, the \textit
{almost surely} condition of (iv) seems somewhat strong, but is still
satisfied by bounded basis functions (e.g., Fourier basis). 

We show in the supplementary material [\citet{PPCAsupp}] that when $\Sig_u$ is diagonal
($u_{it}$ is cross-sectionally independent), both the sub-Gaussian
assumption and condition (iv) can be relaxed.


The following theorem is the main result of this section.

\begin{thmm}\label{th6.1} Under assumptions of Theorem~\ref{th4.1} and
Assumption~\ref{ass6.1}, as $p,T\rightarrow\infty$, if $J$ satisfies
$J=o(\min\{\sqrt{p}, T\})$ and $K<Jd/2$ ($J$ may either grow or stay
constant), we have
\[
P(\widehat K=K)\rightarrow1.
\]
\end{thmm}

\section{Numerical studies}\label{sec7}

This section presents numerical results to demonstrate the performance
of Projected-PCA method for estimating loadings and factors using both
real data and simulated data. 

\subsection{Estimating loading curves with real data}\label{sec7.1}

We collected stocks in S\&P 500 index constituents from CRSP which have
complete daily closing prices from year 2005 through 2013, and their
corresponding market capitalization and book value from Compustat.
There are $337$ stocks in our data set, whose daily excess returns were
calculated.
We considered four characteristics $\bX$ as in \citet{CMO} for each
stock: size, value, momentum and volatility, which were calculated
using the data before a certain data analyzing window so that
characteristics are treated known. See \citet{CMO} for detailed
descriptions of these characteristics. 
All four characteristics are standardized to have mean zero and unit
variance. Note that the construction makes their values independent of
the current data.

\begin{figure}

\includegraphics{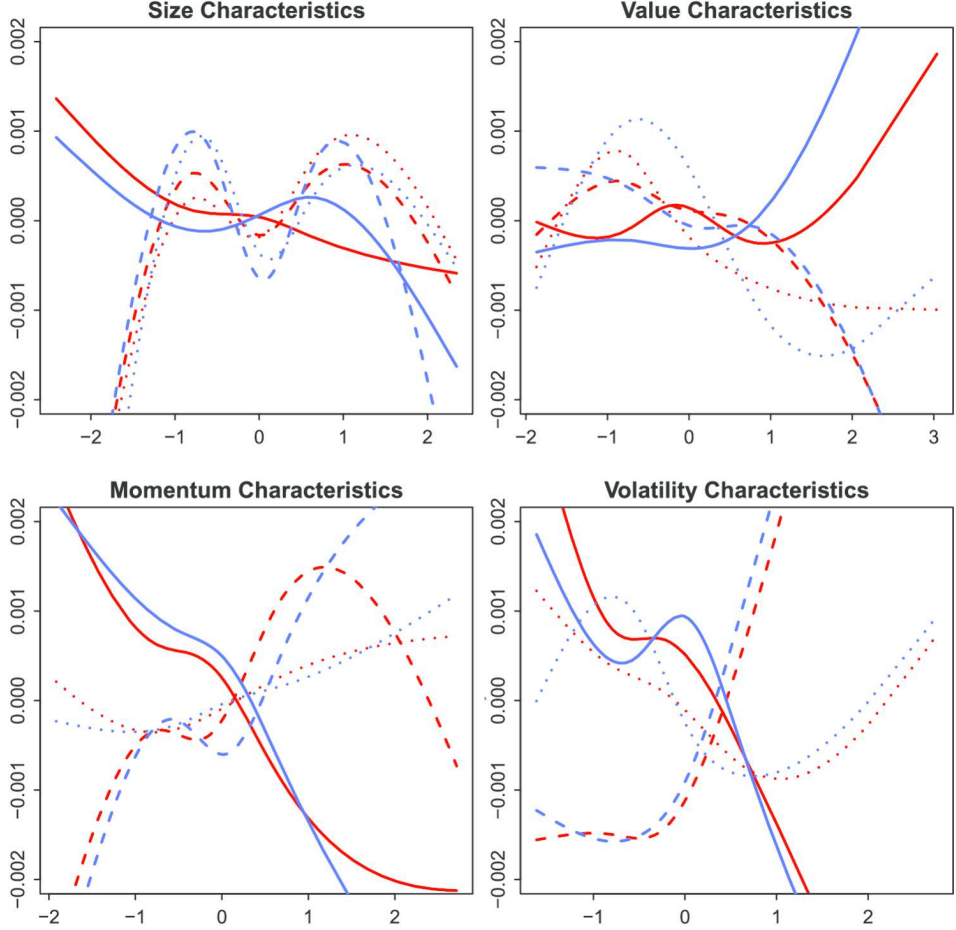}

\caption{Estimated additive loading functions $g_{kl}$, $l=1,\ldots
,4$ from financial returns of 337 stocks in S\&P 500 index. They are
taken as the true functions in the simulation studies.
In each panel (fixed~$l$), the true and estimated curves for $k=1,2,3$
are plotted and compared.
The solid, dashed and dotted red curves are the true curves
corresponding to the first, second and third factors, respectively. The
blue curves are their estimates from one simulation of the calibrated
model with $T=50$, $p=300$.} \label{Fig:Gcurves}
\end{figure}

We fix the time window to be the first quarter of the year 2006, which
contains $T=63$ observations. Given the excess returns $\{ y_{it}\}
_{i\le337, t\le63}$ and characteristics $\bX_i$ as the input data and
setting $K=3$, we fit loading functions $g_k(\bX_i) = \alpha_{ik} +
\sum_{l=1}^4 g_{kl}(X_{il})$ for $k=1, 2, 3$ using the Projected-PCA
method. The four additive components $g_{kl}(\cdot)$ are fitted using
the cubic spline in the R package ``GAM'' with sieve dimension $J=4$.
All the four loading functions for each factor are plotted in
Figure~\ref{Fig:Gcurves}. The contribution of each characteristic to each
factor is quite nonlinear. 

\subsection{Calibrating the model with real data}\label{sec7.2}

We now treat the estimated functions $g_{kl}(\cdot)$ as the true
loading functions, and calibrate a model for simulations. The ``true
model'' is calibrated as follows:
\begin{longlist}[1.]
\item[1.] Take the estimated $g_{kl}(\cdot)$ from the real data as the true
loading functions.
\item[2.] For each $p$, generate $\{\mathbf u_t\}_{t\leq T}$ from $N({\mathbf
0}, \bD
\Sig_0 \bD)$ where $\bD$ is diagonal and $\Sig_0$ sparse. Generate the
diagonal elements of $\bD$ from Gamma($\alpha, \beta$) with $\alpha=
7.06$, $\beta= 536.93$ (calibrated from the real data), and generate
the off-diagonal elements of $\Sig_0$ from $N(\mu_{u}, \sigma_{u}^2)$
with $\mu_{u} = -0.0019$, $\sigma_{u} = 0.1499$. Then truncate $\Sig_0$
by a threshold of correlation $0.03$ to produce a sparse matrix and
make it positive definite by R package ``nearPD.''
\item[3.] Generate $\{\gamma_{ik}\}$ from the i.i.d. Gaussian distribution
with mean $0$ and standard deviation $0.0027$, calibrated with real data.
\item[4.] Generate $\bff_t$ from a stationary VAR model $\bff_t = \bA
\bff
_{t-1} + \bepsilon_t$ where $\bepsilon_t\sim N(\bzero,\Sig_\varepsilon
) $.
The model parameters are calibrated with the market data and listed in
Table~\ref{Table:CalibFactor}.

\item[5.] Finally, generate $\bX_i \sim N(\bzero, \Sig_{X})$. Here $\Sig
_{X}$ is a $4\times4$ correlation matrix estimated from the real data.
\end{longlist}

%
\begin{table}
\caption{Parameters used for the factor generating process, obtained by
calibration to the real data}
\label{Table:CalibFactor}
\begin{tabular*}{\textwidth}{@{\extracolsep{\fill}}lccd{2.4}d{2.4}c@{}}
\hline
\multicolumn{3}{c}{$\bolds{\Sig_\varepsilon}$} &
\multicolumn{3}{c@{}}{$\bolds{\bA}$}
\\
\hline
0.9076 & 0.0049 & 0.0230 & -0.0371 & -0.1226 & $-0.1130$ \\
0.0049 & 0.8737 & 0.0403 & -0.2339 & 0.1060 & $-0.2793$ \\
0.0230 & 0.0403 & 0.9266 & 0.2803 & 0.0755 & $-0.0529$ \\
\hline
\end{tabular*}
\end{table}

\begin{figure}

\includegraphics{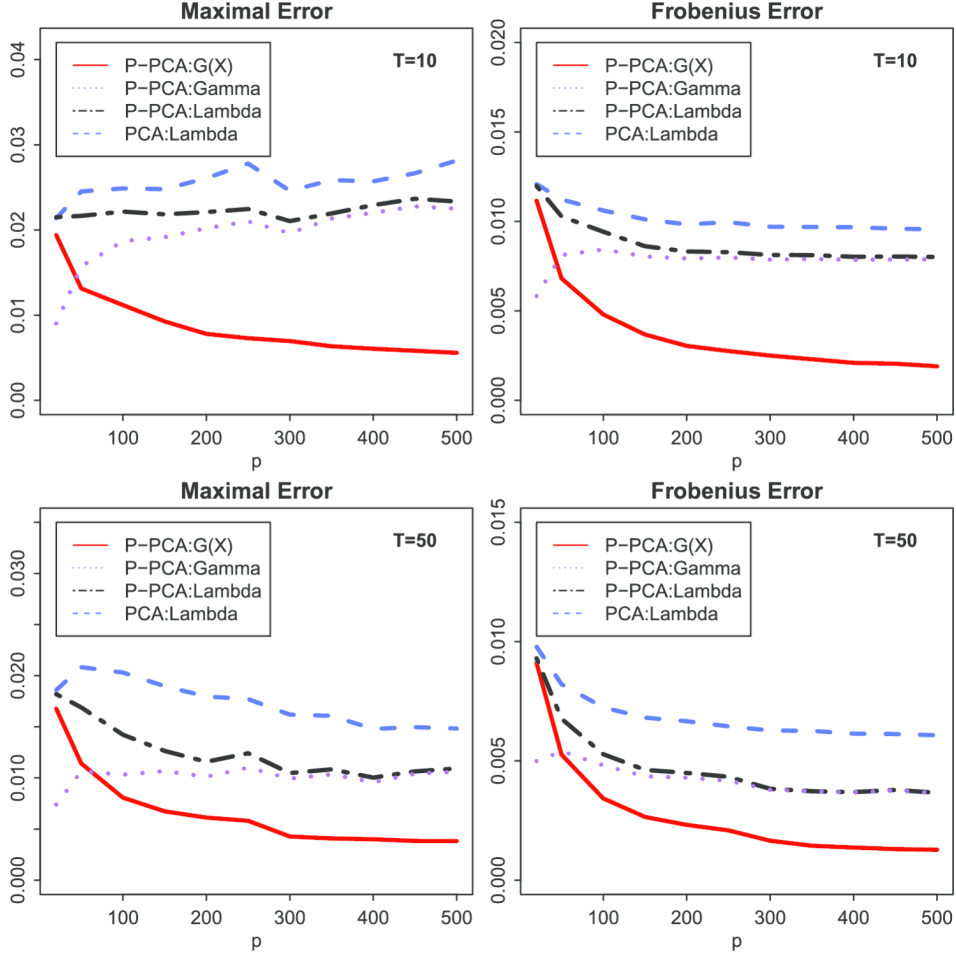}

\caption{Averaged $\| \hat\bLambda- \bLambda\|$ by Projected-PCA
(P-PCA, red solid) and traditional PCA (dashed blue) and $\| \hG- \bG
_0\|$, $\| \hat\bGamma- \bGamma\|$ by P-PCA over 500 repetitions. Left
panel: $ \|\cdot\|_{\max}$, right panel: $\|\cdot\|_F/\sqrt{p}$.} \label{Fig:calibG}
\end{figure}

\begin{figure}

\includegraphics{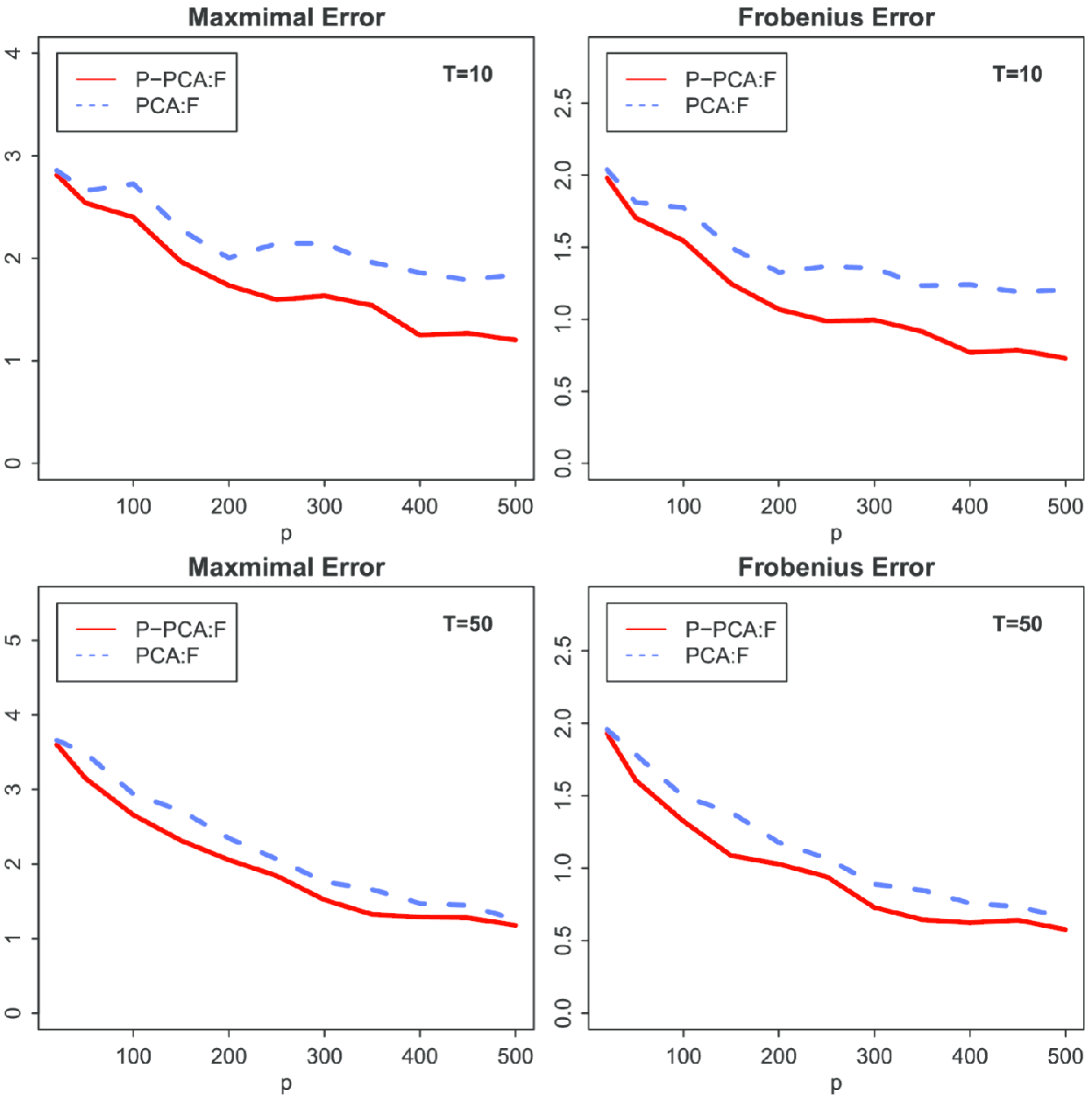}

\caption{Averaged $\|\hF- \bF_0\|_{\max}$ and $\|\hF- \bF_0\|
_{F}/\sqrt{T}$ over 500 repetitions, by Projected-PCA (P-PCA, solid
red) and traditional PCA (dashed blue).} \label{Fig:calibF}
\end{figure}


We simulate the data from the calibrated model, and estimate the
loadings and factors for $T = 10$ and $50$ with $p$ varying from $20$
through $500$.
The ``true'' and estimated loading curves are plotted in Figure~\ref
{Fig:Gcurves} to demonstrate the performance of Projected-PCA. Note
that the ``true'' loading curves in the simulation are taken from the
estimates calibrated using the real data. The estimates based on
simulated data capture the shape of the true curve, though we also
notice slight biases at boundaries. But in general, Projected-PCA fits
the model well.






We also compare our method with the traditional PCA method [e.g., \citet
{SW02}]. The mean values of $\| \hat\bLambda- \bLambda\|_{\max}$,
$\|
\hat\bLambda- \bLambda\|_{F}/\sqrt{p}$, $\|\hF- \bF_0\|_{\max}$ and
$\|\hF- \bF_0\|_{F}/\sqrt{T}$ are plotted in Figures~\ref{Fig:calibG}
and \ref{Fig:calibF}
where $\bLambda= \bG_0(\bX) + \bGamma$ [see Section~\ref{Design2} for
definitions of $\bG_0(\bX)$ and $\bF_0$]. The breakdown error for
$\bG
_0(\bX)$ and $\bGamma$ are also depicted in Figure~\ref{Fig:calibG}. In
comparison, Projected-PCA outperforms PCA in estimating both factors
and loadings including the nonparametric curves $\bG(\bX)$ and random
noise $\bGamma$. The estimation errors for $\bG(\bX)$ of Projected-PCA
decrease as the dimension increases, which is consistent with our
asymptotic theory.

\begin{figure}

\includegraphics{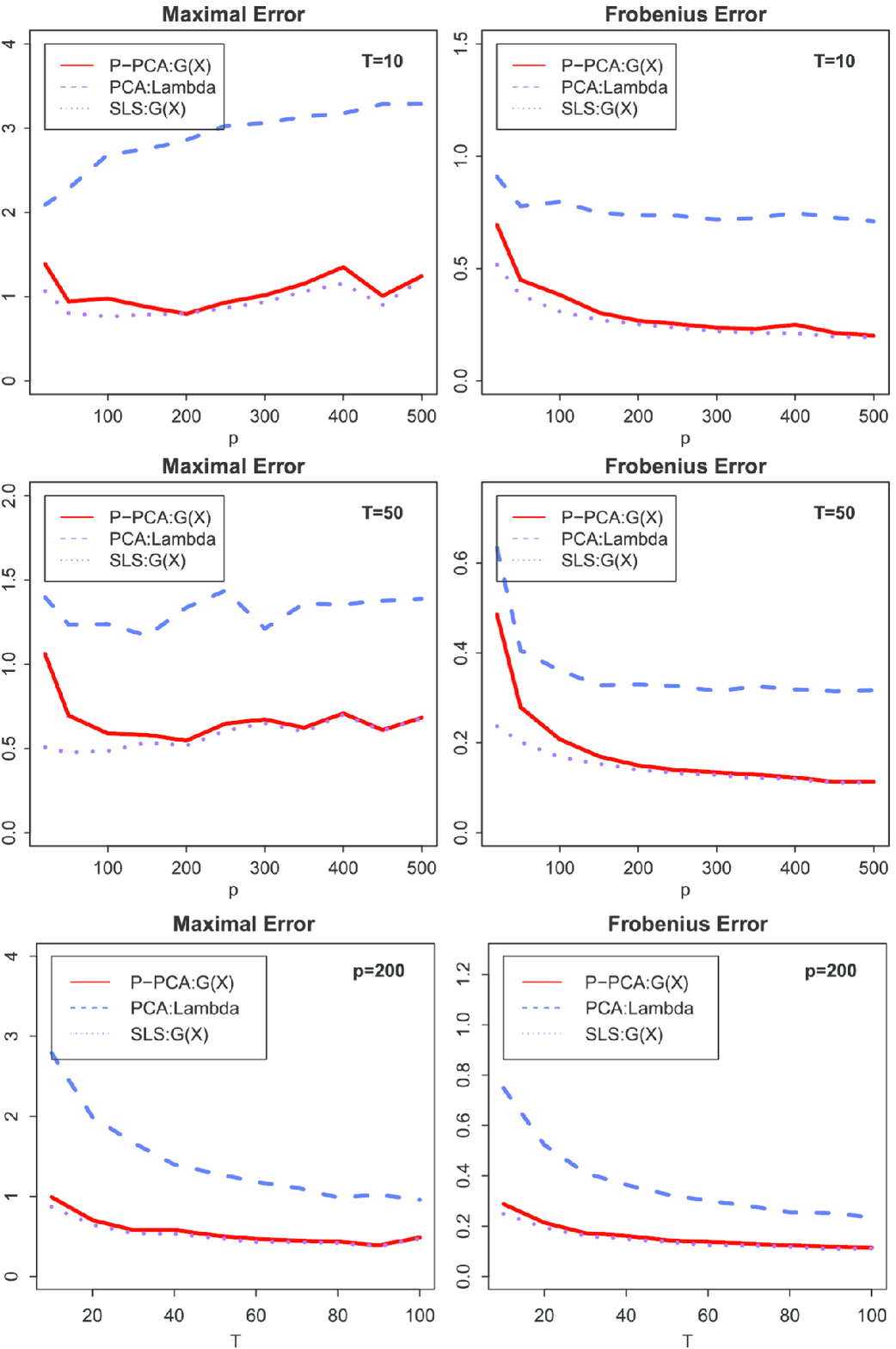}

\caption{Averaged $\| \hG- \bG_0\|_{\max}$ and $\| \hG- \bG_0\|
_F/\sqrt{p}$ over 500 repetitions. P-PCA, PCA and SLS, respectively,
represent Projected-PCA, regular PCA and sieve least squares with known
factors: Design 2. Here, $\bGamma=0$, so $\bLambda=\bG_0$. Upper two
panels: $p$ grows with fixed $T$; bottom panels: $T$ grows with fixed
$p$.} \label{Fig:simpleG}
\end{figure}

\begin{figure}

\includegraphics{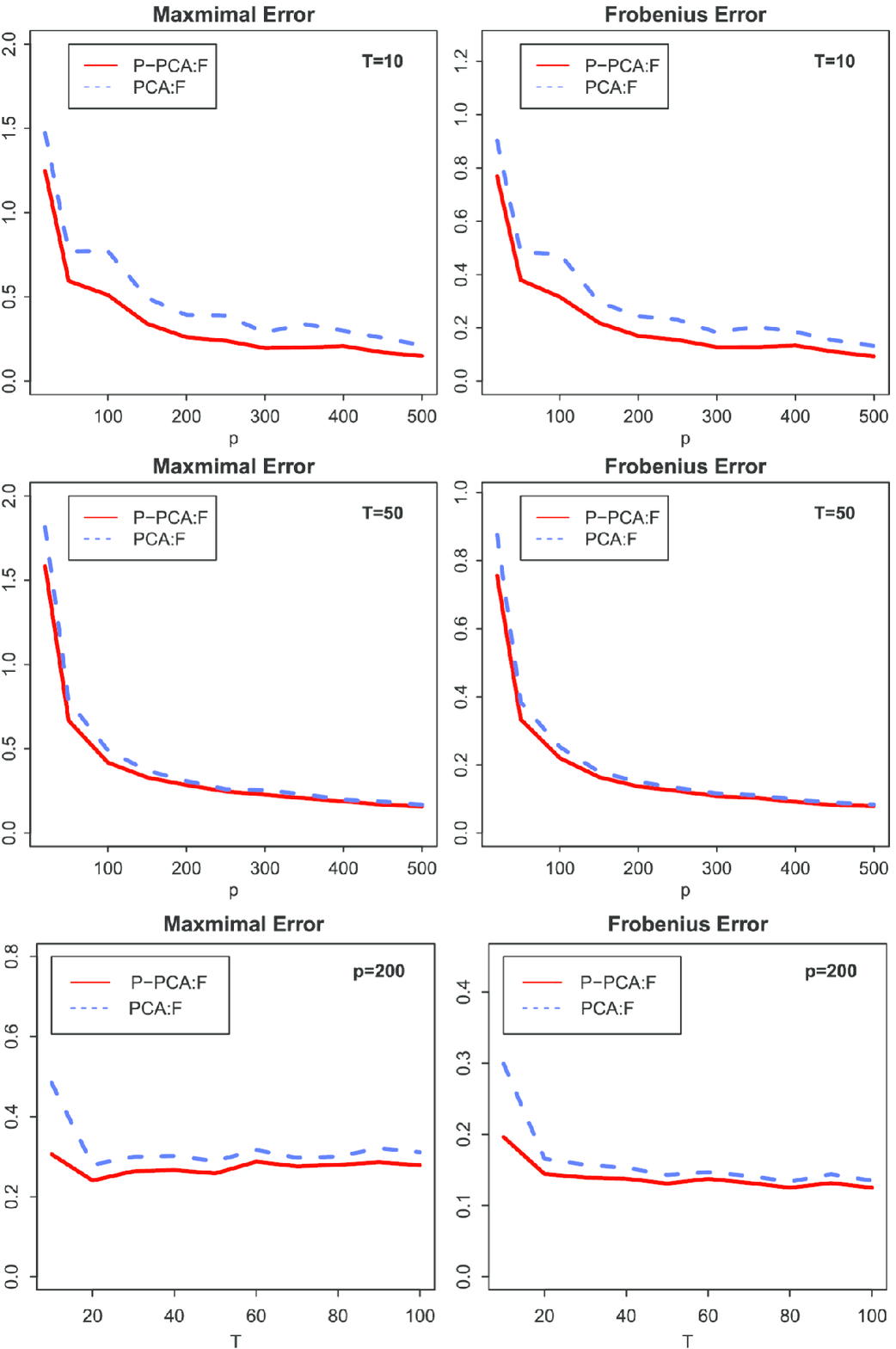}

\caption{Average estimation error of factors over 500 repetitions,
that is, $\|\hF- \bF_0\|_{\max}$ and $\|\hF- \bF_0\|_{F}/\sqrt
{T}$ by
Projected-PCA (solid red) and PCA (dashed blue): Design 2. Upper two
panels: $p$ grows with fixed $T$; bottom panels: $T$ grows with fixed
$p$.} \label{Fig:simpleF}
\end{figure}

\subsection{Design 2} \label{Design2}
Consider a different design with only one observed covariate and three
factors. The three characteristic functions are $g_1 = x, g_2 = x^2-1,
g_3= x^3 - 2x$ with the characteristic $X$ being standard normal.
Generate $\{\bff_t\}_{t\leq T}$ from the stationary VAR(1) model, that
is, $\bff_t = \bA\bff_{t-1} + \bepsilon_t$ where $\bepsilon_t\sim
N(0,\bI)$. We consider $\bGamma= 0$.

We simulate the data for $T = 10$ or $50$ and various $p$ ranging from
$20$ to $500$. To ensure that the true factor and loading satisfy the
identifiability conditions, we calculate a transformation matrix $\bH$
such that $\frac{1}{T} \bH\bF' \bF\bH=\bI_K$, $\bH^{-1} \bG'\bG
\bH
'^{-1}$ is diagonal. Let the final true factors and loadings be $\bF_0
= \bF\bH$, $\bG_0 = \bG\bH'^{-1}$. For each $p$, we run the
simulation for $500$ times.


We estimate the loadings and factors using both Projected-PCA and PC.
For Projected-PCA, as in our theorem, we choose $J=C(p \min
(T,p))^{1/\kappa}$, with $\kappa= 4$ and $C=3$. To estimate the
loading matrix, we also compare with a third method:
sieve-least-squares (SLS), assuming the factors are observable. In this
case, the loading matrix is estimated by $\bP\bY\bF_0/T$, where $\bF_0$
is the true factor matrix of simulated data.

The estimation error measured in max and standardized Frobenius norms
for both loadings and factors are reported in Figures~\ref{Fig:simpleG}
and \ref{Fig:simpleF}. The plots demonstrate the good performance of
Projected-PCA in estimating both loadings and factors. In particular,
it works well when we encounter small $T$ but a large $p$. In this
design, $\bGamma=0$, so the accuracy of estimating $\bLambda= \bG_0$
is significantly improved by using the Projected-PCA.
Figure~\ref{Fig:simpleF} shows that the factors are also better
estimated by Projected-PCA than the traditional one, particularly when
$T$ is small. It is also clearly seen that when $p$ is fixed, the
improvement on estimating factors is not significant as $T$ grows. This
matches with our convergence results for the factor estimator.

It is also interesting to compare Projected-PCA with SLS (Sieve
Least-Squares with observed factors) in estimating the loadings, which
corresponds to the cases of unobserved and observed factors. As we see
from Figure~\ref{Fig:simpleG}, when $p$ is small, the Projected-PCA is
not as good as SLS. But the two methods behave similarly as $p$
increases. This further confirms the theory and intuition that as the
dimension becomes larger, the effects of estimating the unknown factors
are negligible.

\subsection{Estimating number of factors}\label{sec7.4}
We now demonstrate the effectiveness of estimating $K$ by the
projected-PC's eigenvalue-ratio method. The data are simulated in the
same way as in Design 2. $T = 10$ or $50$ and we took the values of $p$
ranging from $20$ to $500$. We compare our Projected-PCA based on the
projected data matrix $\bY'\bP\bY$ to the eigenvalue-ratio test (AH) of
\citet{AH} and \citet{LamYao}, which works on the original data matrix
$\bY'\bY$.

\begin{figure}[b]

\includegraphics{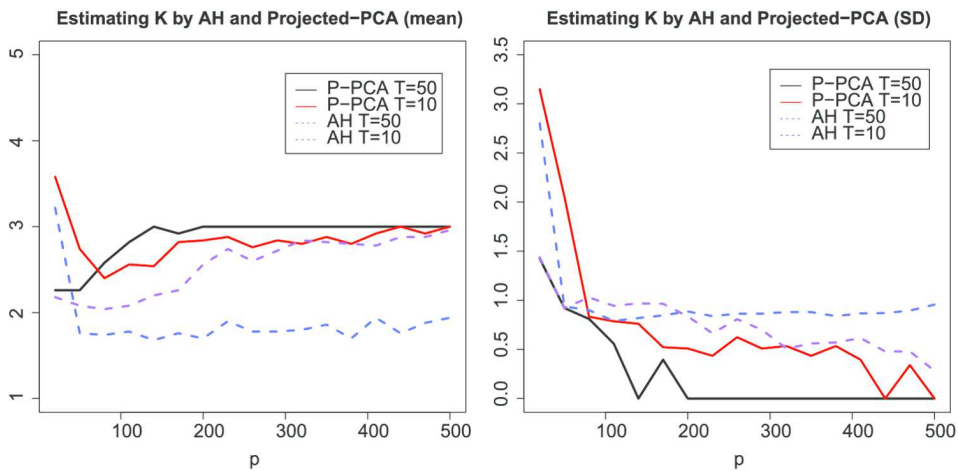}

\caption{Mean and standard deviation of the estimated number of
factors over 50 repetitions. True $K=3$. P-PCA and AH, respectively,
represent the methods of Projected-PCA and \citet{AH}. Left panel: mean;
right panel: standard deviation.} \label{Fig:EstimateK}
\end{figure}

For each pair of $T,p$, we repeat the simulation for $50$ times and
report the mean and standard deviation of the estimated number of
factors in Figure~\ref{Fig:EstimateK}. The Projected-PCA outperforms AH
after projection, which significantly reduces the impact of
idiosyncratic errors. When $T=50$, we can recover the number of factors
almost all the time, especially for large dimensions ($p>200$). On the
other hand, even when $T=10$, projected-PCA still obtains a closer
estimated number of factors. 

\subsection{Loading specification tests with real data}\label{sec7.5}
\begin{figure}

\includegraphics{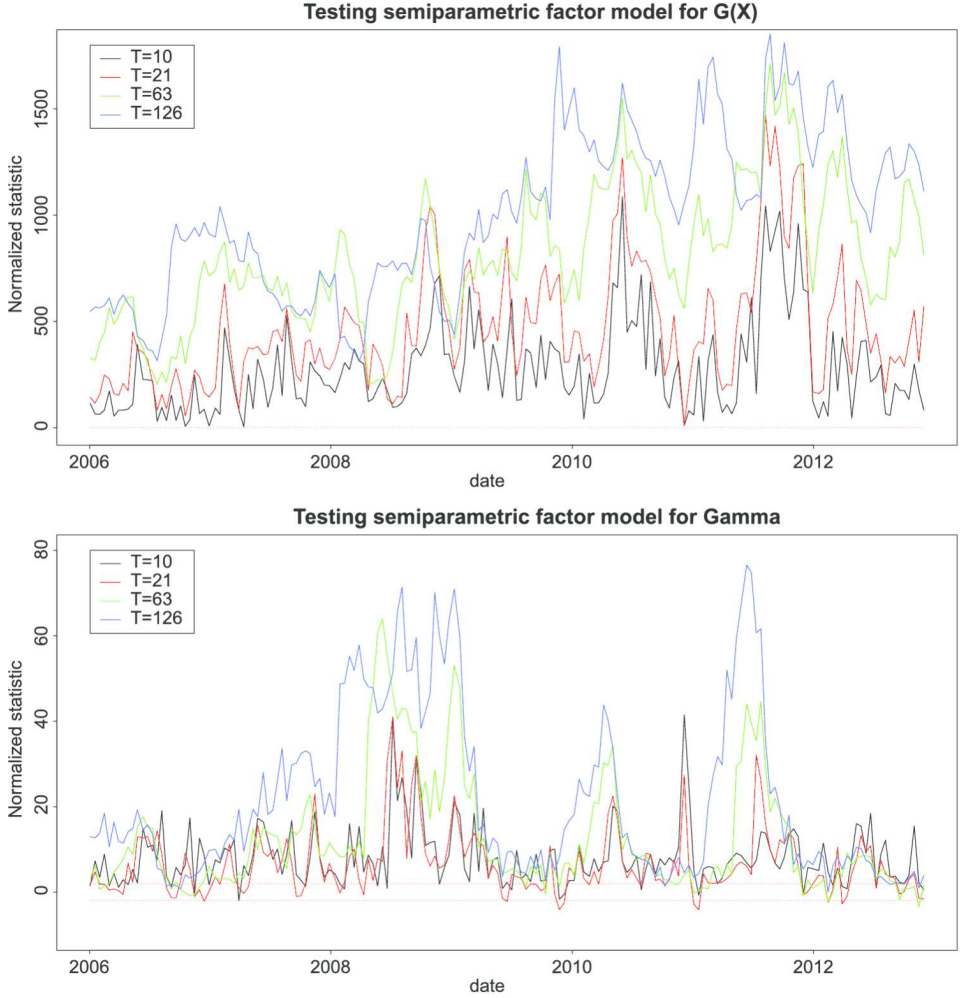}

\caption{Normalized $S_G, S_\Gamma$ from 2006/01/03 to 2012/11/30. The
dotted lines are $\pm1.96$.} \label{Fig:Testing}
\end{figure}

We test the loading specifications on the real data. 
We used the same data set as in Section~\ref{sec7.1}, consisting of excess
returns from 2005 through 2013. The tests were conducted based on
rolling windows, with the length of windows spanning from 10 days, a
month, a quarter and half a year. 
For each fixed window-length ($T$), we computed the standardized test
statistic of $S_G$ and $S_\Gamma$, and plotted them along the rolling
windows respectively in Figure~\ref{Fig:Testing}. 
In almost all cases, the number of factors is estimated to be one in
various combinations of $(T, p, J)$.


Figure~\ref{Fig:Testing} suggests that the semiparametric factor model
is strongly supported by the data. Judging from the upper panel
[testing $H_0^1: \bG(\bX)=0$], we have very strong evidence of the
existence of nonvanishing covariate effect, which demonstrates the
dependence of the market beta's on the covariates $\bX$. In other
words, the market beta's can be explained at least partially by the
characteristics of assets. The results also provide the theoretical
basis for using Projected-PCA to get more accurate estimation.

In the bottom panel of Figure~\ref{Fig:Testing} (testing $H_0^2:
\bGamma
=0$), we see for a majority of periods, the null hypothesis is
rejected. In other words, the characteristics of assets cannot fully
explain the market beta as intuitively expected, and model (\ref
{eq1.2}) in the literature is inadequate. However, fully nonparametric
loadings could be possible in certain time range mostly before
financial crisis. During 2008--2010, the market's behavior had much more
complexities, which causes more rejections of the null hypothesis. The
null hypothesis $\bGamma=0$ is accepted more often since 2012. We also
notice that larger $T$ tends to yield larger statistics in both tests,
as the evidence against the null hypothesis is stronger with larger
$T$. After all, the semiparametric model being considered provides
flexible ways of modeling equity markets and understanding the
nonparametric loading curves. 

\section{Conclusions}\label{sec8}


This paper proposes and studies a high-dimensional factor model with
nonparametric loading functions that depend on a few observed covariate
variables. This model is motivated by the fact that observed variables
can explain partially the factor loadings. We propose a Projected-PCA
to estimate the unknown factors, loadings, and number of factors. After
projecting the response variable onto the sieve space spanned by the
covariates, the Projected-PCA yields a significant improvement on the
rates of convergence than the regular methods. In particular,
consistency can be achieved without a diverging sample size, as long as
the dimensionality grows. This demonstrates that the proposed method is
useful in the typical HDLSS situations. In addition, we propose new
specification tests for the orthogonal decomposition of the loadings,
which fill the gap of the testing literature for semiparametric factor
models. Our empirical findings show that firm characteristics can
explain partially the factor loadings, which provide theoretical basis
for employing Projected-PCA method. On the other hand, our empirical
study also shows that the firm characteristics cannot fully explain the
factor loadings so that the proposed generalized factor model is more
appropriate.

\begin{appendix}\label{app}
\section{Proofs for Section~\texorpdfstring{\protect\ref{1541512515}}{3}}
Throughout the proofs, $p\rightarrow\infty$ and $T$ may either grow
simultaneously with $p$ or stay constant. For two matrices $\bA, \bB$
with fixed dimensions, and a sequence $a_T$, by writing $\bA=\bB
+o_P(a_T)$, we mean $\|\bA-\bB\|_F=o_P(a_T)$.

In the regular factor model $\bY=\bLambda\bF'+\bU$, let $\bK$
denote a
$K\times K$ diagonal matrix of the first $K$ eigenvalues of $\frac
{1}{Tp}\bY'\bP\bY$. Then by definition,
$
\frac{1}{Tp}\bY'\bP\bY\hF=\hF\bK$.
Let $\bM=\frac{1}{Tp}\bLambda'\bP\bLambda\bF'\hF\bK^{-1}$. Then
%
\begin{equation}
\label{ea.1add} \hF-\bF\bM=\sum_{i=1}^3
\bD_i\bK^{-1},
\end{equation}
where
\[
\bD_1=\frac{1}{Tp}\bF\bLambda'\bP\bU\hF,\qquad
\bD_2=\frac
{1}{Tp}\bU'\bP \bU\hF,\qquad
\bD_3=\frac{1}{Tp}\bU'\bP\bLambda
\bF'\hF.
\]

We now describe the structure of the proofs for
\[
\frac{1}{T}\|\hF-\bF\|_F^2 =O_p
\biggl(\frac{J}{p} \biggr).
\]
Note that $\hF-\bF=\hF-\bF\bM+\bF(\bM-\bI)$. Hence, we need to bound
$\frac{1}{T}\|\hF-\bF\bM\|_F^2$ and $\frac{1}{T}\|\bF(\bM-\bI)\|_F^2$,
respectively.

\textit{Step} 1: prove that $\frac{1}{T}\|\hF-\bF\bM\|_F^2=O_P(J/p)$.

Due to the equality (\ref{ea.1add}), it suffices to bound $\|\bK
^{-1}\|
_2$ as well as the $\frac{1}{T}\|\cdot\|_F^2$ norm of $\bD_1, \bD_2, \bD
_3$, respectively. These are obtained in Lemmas \ref{lc.1}, \ref
{lc.2} below.

\textit{Step} 2: prove that $\frac{1}{T}\|\bF'(\hF-\bF\bM)\|
_F=O_P(\sqrt{J/(pT)}+J/p)$.

Still by the equality (\ref{ea.1add}), $\frac{1}{T}\|\bF'(\hF-\bF
\bM)\|
_F\leq\frac{1}{T}\|\bK^{-1}\|_2\sum_{i=1}^3\|\bF'\bD_i\|_F$. Hence,
this step is achieved by bounding $\|\bF'\bD_i\|_F$ for $i=1,2,3$. Note
that in this step, we shall not apply a simple inequality $\|\bF'\bD
_i\|
_F\leq\|\bF\|_F\|\bD_i\|_F$, which is too crude. Instead, with the
help of the result $\frac{1}{T}\|\hF-\bF\bM\|_F^2 =O_p(J/p)$ achieved
in step 1, sharper upper bounds for $\|\bF'\bD_i\|_F$ can be achieved.
We do so in Lemma~B.2 in the supplementary material [\citet{PPCAsupp}].

\textit{Step} 3: prove that $\|\bM-\bI\|_F^2=O_P({J/(pT)}+(J/p)^2)$.

This step is achieved in Lemma~\ref{lc.4} below, which uses the result
in step~2.


Before proceeding to step 1, we first show that the two alternative
definitions for $\widehat\bG(\bX)$ described in Section~\ref{sec2.3}
are equivalent.

\begin{lem}\label{la.1add}
$\frac{1}{T}\bP\bY\widehat\bF=\hXi\widehat\bD^{1/2}$.
\end{lem}

\begin{pf}
Consider the singular value decomposition: $\frac{1}{\sqrt{T}}\bP\bY
=\bV
_1\bS\bV_2'$, where $\bV_1$ is a $p\times p$ orthogonal matrix, whose
columns are the eigenvectors of $\frac{1}{T}\bP\bY\bY'\bP$; $\bV
_2$ is
a $T\times T $ matrix whose columns are the eigenvectors of $ \frac
{1}{T}\bY'\bP\bY$; $\bS$ is a $p\times T$ rectangular diagonal matrix,
with diagonal entries as the square roots of the nonzero eigenvalues
of $\frac{1}{T}\bP\bY\bY'\bP$. In addition, by definition,
$\widehat\bD
$ is a $K\times K$ diagonal matrix consisting of the largest $K$
eigenvalues of $\frac{1}{T}\bP\bY\bY'\bP$; $\hXi$ is a $p\times K$
matrix whose columns are the corresponding eigenvectors. The columns of
$\hF/\sqrt{T}$ are the eigenvectors of $\frac{1}{T}\bY'\bP\bY$,
corresponding to the first $K$ eigenvalues.

With these definitions, we can write $\bV_1=(\widehat\bXi, \tilde
\bV
_1)$, $\bV_2=(\hF/\sqrt{T}, \tilde\bV_2)$, and
\[
\bS=
\pmatrix{ \widehat\bD^{1/2} & \bzero
\vspace*{2pt}
\cr
\bzero& \tilde\bD }
 ,\qquad \hF'
\tilde\bV_2=\bzero, \hF'\hF/T=\bI_K,
\]
for some matrices $\tilde\bV_1, \tilde\bV_2$ and $\tilde\bD$. It then
follows that
\begin{eqnarray*}
\frac{1}{T}\bP\bY\widehat\bF=\bV_1\bS\bV_2'
\frac{1}{\sqrt
{T}}\hF =(\widehat\bXi, \tilde\bV_1)
\pmatrix{ \widehat\bD^{1/2} & \bzero
\cr
\bzero& \tilde\bD }
\pmatrix{\hF'/
\sqrt{T}
\cr
\tilde\bV_2' }
\frac{1}{\sqrt{T}}\hF=\hXi\widehat\bD^{1/2}.
\\
\end{eqnarray*}
\upqed\end{pf}

\begin{lem}\label{lc.1}
$\|\bK\|_2=O_P(1)$, $\|\bK^{-1}\|_2=O_P(1)$, $\|\bM\|_2=O_P(1)$.
\end{lem}

\begin{pf}
The eigenvalues of $\bK$ are the same as those of
\[
\bW=\frac{1}{Tp}\bigl(\Phi(\bX)' \Phi(\bX)
\bigr)^{-1/2}\Phi(\bX)'\bY\bY'\Phi(\bX) \bigl(
\Phi(\bX)'\Phi (\bX)\bigr)^{-1/2}.
\]
Substituting
$\bY=\bLambda\bF'+\bU$,
and $\bF'\bF/T=\bI_K$, we have $\bW=\sum_{i=1}^4\bW_i$, where
\begin{eqnarray*}
\bW_1&=&\frac{1}{p}\bigl(\Phi(\bX)'\Phi(\bX)
\bigr)^{-1/2}\Phi(\bX)' \bLambda \bLambda'
\Phi(\bX) \bigl(\Phi(\bX)'\Phi(\bX)\bigr)^{-1/2},
\\
\bW_2&=&\frac{1}{p}\bigl(\Phi(\bX)'\Phi(\bX)
\bigr)^{-1/2}\Phi(\bX)'\biggl(\frac
{\bLambda\bF'\bU'}{T}\biggr)\Phi(
\bX) \bigl(\Phi(\bX)'\Phi(\bX)\bigr)^{-1/2},
\\
\bW_3&=&\bW_2',
\\
\bW_4&=&
\frac{1}{p}\bigl(\Phi(\bX)'\Phi(\bX)\bigr)^{-1/2}\Phi(
\bX)'\frac
{\bU\bU
'}{T}\Phi(\bX) \bigl(\Phi(\bX)'\Phi(
\bX)\bigr)^{-1/2}. 
\end{eqnarray*}

By Assumption~\ref{ass3.3}, $\|\Phi(\bX)\|_2=\lambda^{1/2}_{\max
}(\Phi
(\bX)'\Phi(\bX))=O_P(\sqrt{p})$,
\begin{eqnarray*}
\bigl\|\bigl( \Phi(\bX)'\Phi(\bX)\bigr)^{-1/2}
\bigr\|_2&=&\lambda^{1/2}_{\max}\bigl(\bigl( \Phi (\bX
)'\Phi(\bX)\bigr)^{-1}\bigr)= O_P
\bigl(p^{-1/2}\bigr),
\\
\|\bP\bLambda\|_2&=&
\lambda_{\max}^{1/2}\biggl(\frac{1}{p}\bLambda'
\bP \bLambda \biggr)p^{1/2}=O_P\bigl(p^{1/2}\bigr).
\end{eqnarray*}
Hence,
\begin{eqnarray*}
\| \bW_2\|_2&\leq&\frac{1}{p}\bigl\|\bigl(\Phi(
\bX)'\Phi(\bX)\bigr)^{-1/2}\bigr\|_2^2\bigl\|
\Phi (\bX)\bigr\|_2\|\bLambda\|_F\biggl\|\frac{1}{T}
\bF'\bU'\Phi(\bX)\biggr\| _F\\
&=&O_P
\biggl(\frac
{1}{pT}\biggr)\bigl\|\bF'\bU'\Phi(\bX)
\bigr\|_F.
\end{eqnarray*}
By Lemma~B.1 in the supplementary material [\citet{PPCAsupp}],
$
\|\bW_2\|_2= O_P(\frac{\sqrt{J}}{\sqrt{pT}})
$. Similarly,
\begin{eqnarray*}
\|\bW_4\|_2&\leq&\frac{1}{pT}\bigl\|\bigl(\Phi(
\bX)'\Phi(\bX)\bigr)^{-1/2}\bigr\|_2^2\bigl\|
\Phi (\bX)'\bU\bigr\|_F^2\\
&= &O_P
\biggl( \frac{1}{p^2T}\biggr)\bigl\|\Phi(\bX)'\bU\bigr\|
_F^2=O_P\biggl(\frac{J}{p}\biggr).
\end{eqnarray*}
Using the inequality that for the $k$th eigenvalue, $|\lambda_k(\bW
)-\lambda_k(\bW_1)|\leq\|\bW-\bW_1\|_2$,
we have $|\lambda_k(\bW)-\lambda_k(\bW_1)|=O_P(T^{-1/2}+p^{-1})$, for
$k=1,\ldots,K$.
Hence, it suffices to prove that the first $K$ eigenvalues of $\bW_1$
are bounded away from both zero and infinity, which are also the first
$K $ eigenvalues of
$\frac{1}{p}\bLambda'\bP\bLambda$. This holds under the theorem's
assumption (Assumption~\ref{ass3.1}). Thus, $\|\bK^{-1}\|_2 =
O_P(1)=\|
\bK\|_2$, which also implies $\|\bM\|_2=O_P(1)$.
\end{pf}

\begin{lem}\label{lc.2}
\textup{(i)} $\|\bD_1\|_F^2=O_P(TJ/p)$, \textup{(ii)}
$\|\bD_2\|_F^2=O_P(J/p^2)$,
\textup{(iii)}\break $\|\bD_3\|_F^2=O_P(TJ/p)$,
\textup{(iv)} $\frac{1}{T}\|\hF-\bF\bM\|_F^2=O_P(J/p)$.
\end{lem}

\begin{pf} It follows from Lemma~B.1 in the supplementary
material [\citet{PPCAsupp}] that $\|\bP\bU\|_F=O_P(\sqrt{TJ})$. Also, $\|\bF\|
_F^2=O_P(T)=\|\hF\|_F^2$ and Assumption~\ref{ass3.1} implies $\|\bP
\bLambda\|_2^2=O_P(p)$. So
\begin{eqnarray*}
\|\bD_1\|_F^2&=&\biggl\|\frac{1}{Tp}\bF
\bLambda'\bP\bU\hF\biggr\|_F^2\leq
\frac
{1}{T^2p^2}\|\bF\|_F^2\|\hF\|_F^2
\|\bP\bLambda\|_2^2\|\bP\bU\| _F^2=O_P(TJ/p),
\\
\| \bD_2\|_F^2&=&\biggl\|\frac{1}{Tp}
\bU'\bP\bU\hF\biggr\|_F^2\leq\frac
{1}{T^2p^2}\|
\bP\bU\|_F^2\|\hF\|_F^2=O_P
\bigl(J/p^2\bigr),
\\
\|\bD_3\|_F^2&=&
\biggl\|\frac{1}{Tp}\bU'\bP\bLambda\bF'\hF
\biggr\|_F^2\leq \frac
{1}{T^2p^2}\|\bP\bU
\|_F^2\|\bP\bLambda\|_2^2\|\bF
\|_F^2\|\hF\| _F^2=
O_P(TJ/p).
\end{eqnarray*}

By Lemma~\ref{lc.1}, $\|\bK^{-1}\|_2=O_P(1)$. Part (iv) then follows
directly from
\[
\frac{1}{T}\|\hF-\bF\bM\|_F^2\leq
O_P\biggl(\frac{1}{T}\bigl\|\bK^{-1}\bigr\| _2
\biggr) \bigl(\|\bD _1\|_F^2+\|
\bD_2\|_F^2+\|\bD_3
\|_F^2\bigr).
\]
\upqed\end{pf}

\begin{lem} \label{lc.4} In the regular factor model, $\|\bM-\bI\|
_F=O_P(\sqrt{J/(pT)}+J/p)$.
\end{lem}

\begin{pf}
By Lemma~B.2 in the supplementary material [\citet{PPCAsupp}] and the
triangular inequality, $\|\frac{1}{T}(\hF-\bF\bM)'\bF\|=O_P(\sqrt
{J/(pT)}+J/p)$. Hence,
\[
\hF'\bF/T=\bM'+\frac{1}{T}(\hF-\bF
\bM)'\bF=\bM'+O_P\bigl(\sqrt {J/(pT)}+J/p
\bigr).
\]
Right multiplying $\bM$ to both sides $\hF'\bF\bM/T=\bM'\bM
+O_P(\sqrt
{J/(pT)}+J/p)$. In addition,
\begin{eqnarray*}
\bigl\|\hF'(\hF-\bF\bM)/T\bigr\|_F&\leq&\frac{1}{T}\|\hF-
\bF\bM\|_F^2+\bigl\| \bF'(\hF -\bF\bM)/T
\bigr\|_F\\
&=&O_P\bigl(\sqrt{J/(pT)}+J/p\bigr).
\end{eqnarray*}

Hence,
\[
\bI=\bM'\bM+O_P\bigl(\sqrt{J/(pT)}+J/p\bigr).
\]
In addition, from $\bM=\frac{1}{Tp}\bLambda'\bP\bLambda\bF'\hF
\bK
^{-1}=\frac{1}{p}\bLambda'\bP\bLambda\bM\bK^{-1}+O_P(\sqrt
{J/(pT)}+J/p)$,
\[
\bM\bK=\frac{1}{p}\bLambda'\bP\bLambda\bM+O_P
\bigl(\sqrt{J/(pT)}+J/p\bigr).
\]
Because $\bLambda'\bP\bLambda$ is diagonal, the same proofs of those of
Proposition~C.3 lead to the desired result.
\end{pf}



\begin{pf*}{Proof of Theorem~\ref{th3.1}}
It follows from Lemmas \ref{lc.2}(iv) and \ref{lc.4} that
\[
\frac{1}{T}\|\hF-\bF\|_F^2\leq\frac{2}{T}
\|\hF-\bF\bM\|_F^2+2\| \bM-\bI \|_F^2=O_p
\biggl(\frac{J}{p} \biggr).
\]
As for the estimated loading matrix, note that
\[
\hG(\bX)=\frac{1}{T}\bP\bY\hF=\frac{1}{T}\bP\bLambda
\bF'\hF +\frac
{1}{T}\bP\bU\hF=\bP\bLambda+\bE,
\]
where $\bE=\frac{1}{T}\bP\bLambda\bF'(\hF-\bF)+\frac{1}{T}\bP
\bU(\hF-\bF
)+\frac{1}{T}\bP\bU\bF$.

By Lemmas B.2 and \ref{lc.4},
\begin{eqnarray*}
\biggl\|\frac{1}{T}\bP\bLambda\bF'(\hF-\bF)\biggr\|_F&\leq&
O_P \biggl(\frac
{\sqrt
{p}}{T} \biggr)\bigl\|\bF'(\hF-\bF
\bM)\bigr\|_F+O_P(\sqrt{p})\|\bM-\bI\| _F\\
&=&O_P
\biggl(\sqrt{\frac{J}{T}}+\frac{J}{\sqrt{p}} \biggr).
\end{eqnarray*}
By Lemma~B.1, $
\|\frac{1}{T}\bP\bU(\hF-\bF)\|_F\leq\frac{1}{T}\|\bP\bU\|_2\|
\hF-\bF\|
_F=O_P( \frac{J}{\sqrt{p}} )$, and from Lemma~B.2 $\|\frac
{1}{T}\bP\bU\bF\|_F = O_P(\sqrt
{\frac{J}{T}})$.
Hence, $\|\bE\|_F=O_P(\sqrt{\frac{J}{T}}+\frac{J}{\sqrt{p}})$,
which implies
\[
\frac{1}{p}\bigl\|\hG(\bX)-\bP\bLambda\bigr\|_F^2=
O_P \biggl(\frac
{J}{pT}+\frac
{J^2}{p^2} \biggr).
\]
%
\upqed\end{pf*}

All the remaining proofs are given in the supplementary material [\citet{PPCAsupp}].
\end{appendix}

\begin{supplement}[id=suppA]
\stitle{Technical proofs \citet{PPCAsupp}}
\slink[doi]{10.1214/15-AOS1364SUPP} 
\sdatatype{.pdf}
\sfilename{aos1364\_supp.pdf}
\sdescription{This supplementary material contains all the remaining proofs.}
\end{supplement}


%
%


%
%

%



\printaddresses
\end{document}